\documentclass[a4paper,fleqn,usenatbib,usedcolumn,useAMS]{mnras}

\usepackage{mathptmx}
\usepackage{txfonts}

\usepackage[T1]{fontenc}
\usepackage{ae,aecompl}

\usepackage{graphicx}	
\usepackage{subfig}
\usepackage{lscape}
\usepackage{supertabular,booktabs}
\usepackage{longtable,lscape}
\usepackage{rotating}
\usepackage{multirow,multicol}

\def\reff@jnl#1{{\rm#1\/}}

\def\aj{\reff@jnl{AJ}}                  
\def\araa{\reff@jnl{ARA\&A}}            
\def\apj{\reff@jnl{ApJ}}                        
\def\apjl{\reff@jnl{ApJ}}               
\def\apjs{\reff@jnl{ApJS}}              
\def\ao{\reff@jnl{Appl.Optics}}         
\def\apss{\reff@jnl{Ap\&SS}}            
\def\aap{\reff@jnl{A\&A}}               
\def\aapr{\reff@jnl{A\&A~Rev.}}         
\def\aaps{\reff@jnl{A\&AS}}             
\def\azh{\reff@jnl{AZh}}                        
\def\baas{\reff@jnl{BAAS}}              
\def\jrasc{\reff@jnl{JRASC}}            
\def\memras{\reff@jnl{MmRAS}}           
\def\mnras{\reff@jnl{MNRAS}}            
\def\pra{\reff@jnl{Phys. Rev. A}}         
\def\prb{\reff@jnl{Phys. Rev. B}}         
\def\prc{\reff@jnl{Phys. Rev. C}}         
\def\prd{\reff@jnl{Phys. Rev. D}}         
\def\prl{\reff@jnl{Phys. Rev. Lett}}      
\def\pasp{\reff@jnl{PASP}}              
\def\pasj{\reff@jnl{PASJ}}              
\def\qjras{\reff@jnl{QJRAS}}            
\def\rmxaa{\reff@jnl{RMxAA}}		
\def\skytel{\reff@jnl{S\&T}}            
\def\solphys{\reff@jnl{Solar~Phys.}}    
\def\sovast{\reff@jnl{Soviet~Ast.}}     
\def\ssr{\reff@jnl{Space~Sci.Rev.}}     
\def\zap{\reff@jnl{ZAp}}                        
\def\nat{\reff@jnl{Nature}}             

\def\p#1by#2{{\partial{#1} \over \partial{#2}}}
\def\pp#1by#2#3{{\partial^2{#1} \over \partial{#2}\partial{#3}}}
\def\d#1by#2{{{\rm d}{#1} \over {\rm d}{#2}}}
\def\dd#1by#2#3{{{\rm d}^2{#1} \over {\rm d}{#2}{\rm d}{#3}}}

\title[Taurus with GMRT]{A GMRT survey of regions towards the Taurus Molecular Cloud at 323 and 608\,MHz}

\author[R. E. Ainsworth et~al.]{
Rachael E. Ainsworth,$^{1}$\thanks{E-mail: rainsworth@cp.dias.ie}
Colm P. Coughlan,$^{1}$
David A. Green,$^{2}$
Anna M. M. Scaife,$^{3}$
\newauthor and Tom P. Ray$^{1}$
\\
$^{1}$Dublin Institute for Advanced Studies, School of Cosmic Physics, 31 Fitzwilliam Place, Dublin D02 XF86, Ireland\\
$^{2}$Astrophysics Group, Cavendish Laboratory, J J Thomson Avenue, Cambridge CB3 0HE, UK\\
$^{3}$Jodrell Bank Centre for Astrophysics, School of Physics and Astronomy, The University of Manchester, Oxford Road, Manchester M13 9PL, UK\\
}

\date{Accepted XXX. Received YYY; in original form ZZZ}

\pubyear{2015}

\begin{document}
\label{firstpage}
\pagerange{\pageref{firstpage}--\pageref{lastpage}}
\maketitle

\begin{abstract}
We present observations of three active sites of star formation in the Taurus Molecular Cloud complex taken at 323 and 608\,MHz (90 and 50\,cm, respectively) with the Giant Metrewave Radio Telescope (GMRT). Three pointings were observed as part of a pathfinder project, targeted at the young stellar objects (YSOs) L1551~IRS~5, T~Tau and DG~Tau (the results for these target sources were presented in a previous paper). In this paper, we search for other YSOs and present a survey comprising of all three fields; a by-product of the large instantaneous field of view of the GMRT. The resolution of the survey is of order 10\,arcsec and the best rms noise at the centre of each pointing is of order $100\,\umu$Jy\,beam$^{-1}$ at 323\,MHz and $50\,\umu$Jy\,beam$^{-1}$ at 608\,MHz. We present a catalogue of 1815 and 687 field sources detected above $5\,\sigma_{\rm rms}$ at 323 and 608\,MHz, respectively. A total of 440 sources were detected at both frequencies, corresponding to a total unique source count of 2062 sources. We compare the results with previous surveys and showcase a sample of extended extragalactic objects. Although no further YSOs were detected in addition to the target YSOs based on our source finding criteria, these data can be useful for targeted manual searches, studies of radio galaxies or to assist in the calibration of future observations with the Low Frequency Array (LOFAR) towards these regions. 
\end{abstract}




\section{Introduction}
\label{intro}

\begin{figure*}
\includegraphics[width=0.7\textwidth]{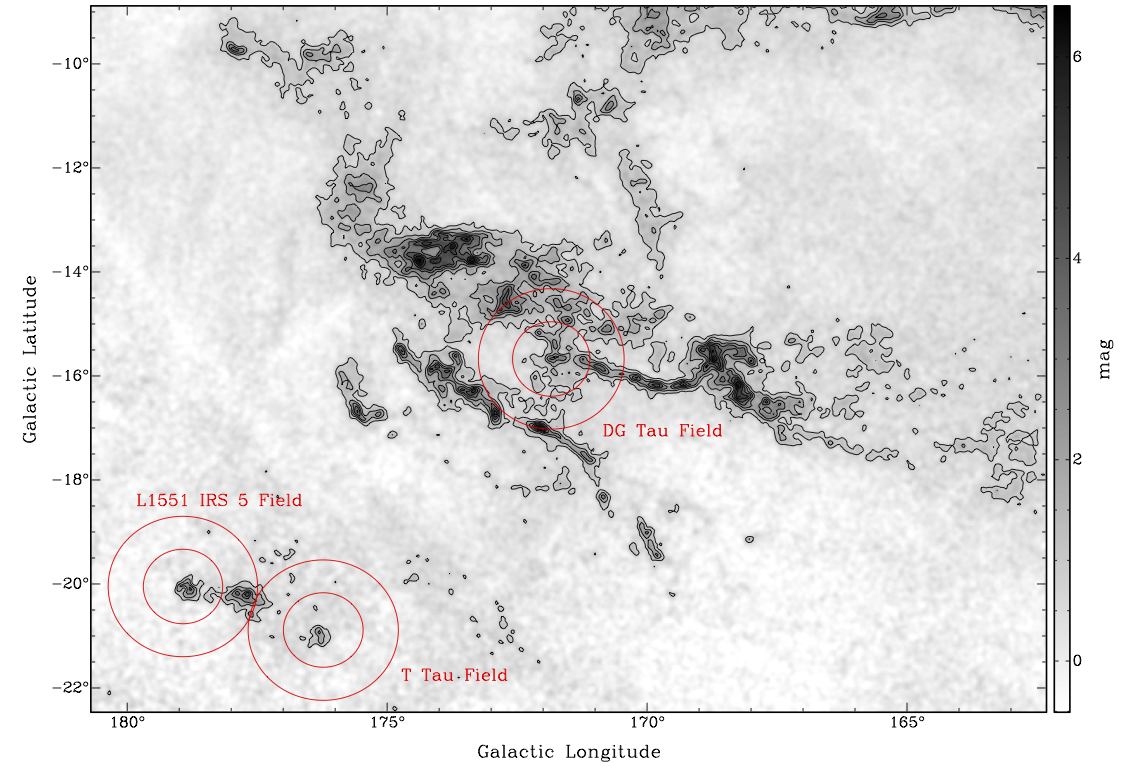}
\caption[The Taurus Molecular Cloud]{Overview of the TMC. Greyscale is the high resolution CO extinction map from \citet{2005PASJ...57S...1D}. Circles correspond to the GMRT observed fields (the FWHM of the GMRT primary beam is $\approx85$\,arcmin at 323\,MHz and $\approx44$\,arcmin at 608\,MHz). Axes are J2000.0 Galactic coordinates.}
\label{fig:taurus}
\end{figure*}

The Taurus Molecular Cloud (TMC, see Fig.~\ref{fig:taurus}) is one of the best studied star forming regions due to its relatively nearby rich population of low mass young stellar objects \citep[YSOs;][]{2008hsf1.book..405K}. The TMC is located at a mean distance of 140\,pc. However, recent observations with the Very Long Baseline Array have been used to refine distance measurements and determine the three-dimensional structure of the complex \citep[which has been shown to have a depth of 30\,pc;][]{2005ApJ...619L.179L, 2009ApJ...698..242T}. The TMC is not as densely populated as other star forming regions, such as $\rho$~Ophiuchus and Orion, which allows for the study of individual low-mass protostellar systems analogous to the Sun. The low stellar density also minimises the mutual influence of outflows, jets or gravitational effects on star formation and the lack of more luminous stars (there are no O stars and only very few B and A stars) limits the effects of strong stellar winds and ionising UV radiation \citep{2007A&A...468..353G}. For these reasons, the TMC allows tests of stellar evolution models and provides the standard initial mass function for a nearby young association with stellar ages of $\sim1$--10\,Myr \citep[e.g.][]{1995ApJS..101..117K, 2004ApJ...617.1216L}. 

The TMC has been extensively surveyed in most bands of the electromagnetic spectrum, from X-ray, infrared and optical \citep[see e.g.][]{2007prpl.conf..329G} to sub-millimeter \citep{2005ApJ...631.1134A} wavelengths, providing an unsurpassed database for the nearest major star forming cloud complex. There have not, however, been many large-scale surveys at radio (centimetre) wavelengths directed at the TMC. \citet{2015ApJ...801...91D} aimed to rectify this deficit by conducting a multi-epoch radio study of the TMC at 4.5 and 7.5\,GHz as part of the Gould's Belt Very Large Array Survey (GBS-VLA) to systematically characterise the centimetre-wave properties of the YSO population. This region was also covered by all sky surveys such as the NRAO VLA Sky Survey \citep[NVSS,][]{1998AJ....115.1693C} at 1.4\,GHz and the Low Frequency Array \citep[LOFAR,][]{2013A&A...556A...2V} Multifrequency Snapshot Sky Survey \citep[MSSS,][]{2015A&A...582A.123H} at 30--160\,MHz, albeit with relatively poor angular resolution and sensitivity. There are no published surveys targeted at YSOs in the TMC at frequencies $<1$\,GHz. Furthermore, the study of YSOs at radio wavelengths has previously been largely confined to frequencies $\nu>1$\,GHz. This is due to the past sensitivity limitations of radio telescopes, the radio weakness of YSOs (flux densities of order $\sim1$\,mJy at centimetre wavelengths), the fact that they are typically detected via thermal bremsstrahlung (free--free) radiation and thus their spectra rise with frequency. 

We conducted a pathfinder project with the Giant Metrewave Radio Telescope (GMRT) to extend the study of young stars to very low radio frequencies. We observed the well-studied TMC members L1551~IRS~5, T~Tau and DG~Tau at frequencies $\nu=323$ and 608\,MHz (90 and 50\,cm, respectively), the results of which were presented in \citet{2016MNRAS.459.1248A}. A natural by-product of these GMRT observations was a large instantaneous field of view within which to search for additional objects, in particular other YSOs. Although the TMC is not as densely populated with young stars as other star forming regions, there are several other known pre-main-sequence objects located within the GMRT field of view of each of the target fields.  For example, in the southern region of the TMC, most pre-main sequence stars are in and around the Lynds~1551 (L1551) dark cloud. In addition to L1551~IRS~5, the protostar L1551~NE and a few deeply embedded T~Tauri stars (HL~Tau, XZ~Tau and HH~30~IRS) form a close group of pre-main sequence stars \citep{2008hsf1.book..405K}. The entirety of the L1551 cloud fits within the 44\,arcmin half-power point of the GMRT primary beam at 608\,MHz (see Fig.~\ref{fig:taurus}) and therefore the GMRT can be a potentially useful survey instrument for star forming regions at very long wavelengths due to the extent of its field of view. 

The importance of investigating the radio emission from young stars at very long wavelengths is twofold. First, observations of the long wavelength turnover in the free-free spectrum can constrain physical properties of the ionised plasma from these systems such as gas mass and electron density \citep[see e.g.][]{2016MNRAS.459.1248A}. Second, the spectra of non-thermal emission processes \citep[such as those observed from e.g.][]{2010Sci...330.1209C, 2014ApJ...792L..18A} typically rise at longer wavelengths making them easier to detect. It is important to note however, that non-thermal coronal emission, which has been detected from a large fraction of YSOs as part of the GBS-VLA \citep[e.g.][]{2013ApJ...775...63D, 2015ApJ...801...91D} can exhibit a broad range of spectral indices and may therefore be a poor discriminant between thermal and non-thermal emission.

The new generation of radio interferometers such as LOFAR and upgrades to existing facilities such as the GMRT, will allow access to a previously unexplored wavelength regime for young stars. GMRT data can also be used to assist in the calibration of observations with LOFAR towards these regions as the radio sky at such low frequencies is largely unknown in these directions. 

In this paper we present a full catalogue of sources detected within the GMRT field of view of each target field, including a detailed description of the survey methodology and data products. In Section~\ref{sec:obs} we provide details of the observations and data reduction. In Section~\ref{sec:cat_creation} we discuss the method used for source fitting, measuring spectral indices, catalogue creation and a sample table of the final catalogue. We also showcase a sample of extended sources. In Section~\ref{sec:dis} we discuss the YSO detections and notable non-detections. We compare between the 323 and 608\,MHz images and with previous surveys. In Section~\ref{sec:dp} we list the resulting data products and summarise the survey in Section~\ref{sec:summary}.

\vspace{-10pt}
\section{Observations and Data Reduction}
\label{sec:obs}

\begin{table*}
\centering
\caption{Observing details. Column [1] contains the target source name at the phase centre of the field; [2] the Right Ascension of the phase centre; [3] the Declination of the phase centre; [4] the on-source observing time; [5] the central frequency of the final images (post-processing); [6] the dimensions of the native \textsc{clean} restoring beam of the 323\,MHz images of \citet{2016MNRAS.459.1248A} used to re-image both the 323 and 608\,MHz data to have matching beam and cell sizes for source fitting in \textsc{PyBDSM}; and [7] the rms noise of a local patch of sky at the centre of the field. See Section~\ref{sec:obs} for details.}
\label{tab:srclist}
\begin{tabular}{lcccccc} 
\hline
Field & \multicolumn{2}{c}{J$2000.0$ Coordinates} & Obs. time & $\nu$ & FWHM, PA & $\sigma_{\rm{rms}}$  \\
       & $\alpha$ ($^{\rm h}~~^{\rm m}~~^{\rm s}$) & $\delta$ ($^{\circ}$~$'$~$''$) & (hrs.) & (MHz) & ($''\times''$, $^\circ$) & ($\umu$Jy\,beam$^{-1}$)  \\
\hline
L1551~IRS~5 & 04~31~34.1 & +18~08~04.8 & $6.0$ & $323$ & $11.4\times9.5$, $-88.5$ & 157  \\
	& & & $2.2$ & $608$ & $11.4\times9.5$, $-88.5$ & 54 \\
T~Tau & 04~21~59.4 & +19~32~06.4 & $3.3$ & $323$ & $10.8\times9.5$, $-81.6$ & 98 \\
	 & & & $2.2$ & $608$ & $10.8\times9.5$, $-81.6$ & 50 \\
DG~Tau & 04~27~04.7 & +26~06~16.3 & $6.0$ & $323$ & $11.6\times9.2$, $79.6$ & 141 \\
	& & & $2.2$ & $608$ & $11.6\times9.2$, $79.6$ & 90 \\
\hline
\end{tabular}
\end{table*}

The details of the observations and data reduction using the Astronomical Image Processing Software (\textsc{AIPS}) were presented in \citet{2016MNRAS.459.1248A} which we reiterate here for the convenience of the reader. Observations centred on the young stars L1551~IRS~5, T~Tau and DG~Tau were made with the GMRT \citep[see e.g.][]{2005ICRC...10..125A} in 325 and 610\,MHz observing modes between 2012 December 6--14 (average epoch 2012.95). The GMRT comprises thirty 45\,m dishes, however an average of 27 antennas were operational during each observing run. At 325\,MHz, observations were taken for 7\,hours per night over the course of three nights for a total of 21\,hours and observations at 610\,MHz were taken for 10\,hours in a single run (see Table~\ref{tab:srclist} for the number of hours on-source for each individual field). A total bandwidth of 32\,MHz was observed, which was split into 256 spectral channels. The sample integration time was 16.9\,s. 

Observations of 3C48, 3C147 or 3C286 were made at the beginning and end of each observing run to calibrate the flux density scale. The flux densities were calculated using the task \textsc{setjy} and were found to be 45.6\,Jy at 325\,MHz and 29.4\,Jy at 610\,MHz for 3C48, 55.1\,Jy at 325\,MHz for 3C147, and 20.8\,Jy at 610\,MHz for 3C286 using the \citet{2013ApJS..204...19P} scale. Each target field was observed for a series of interleaved 10\,min scans so as to maximise the \textit{uv}~coverage for imaging. The nearby phase calibrator, J0431+206, was observed for 3\,min after every two scans to monitor the phase and amplitude fluctuations of the telescope. The task \textsc{getjy} retrieved flux densities of $2.78\pm0.02$\,Jy at 325\,MHz and $3.05\pm0.05$\,Jy at 610\,MHz for J0431+206.
 
Flagging of baselines, antennas, channels and scans that suffered heavily from interference was performed for each observing run using standard \textsc{AIPS} tasks. Bandpass calibration was applied to each antenna using the flux calibrator sources. Ten central frequency channels were then combined together with the task \textsc{splat} and antenna-based phase and amplitude calibration was performed with \textsc{calib}. This calibration was then applied to the full 256 channel dataset and averaged into 24 separate spectral channels. Some end channels, where the bandpass correction is larger, were omitted. The effective mean frequency of the observations was therefore 322.665 and 607.667\,MHz (hereafter referred to as 323 and 608\,MHz, respectively throughout this paper) with an effective bandwidth of 30\,MHz covered by the averaged 24 channels. Target source data was then extracted from the larger datasets via the \textsc{split} task and the data from each observing run were concatenated with \textsc{dbcon} for self-calibration and imaging. 

Imaging the large field of view of the GMRT directly leads to significant phase errors due to the non-planar nature of the sky \citep[see e.g.][]{2007MNRAS.376.1251G}. To minimise these errors, we conducted wide-field imaging using facets following \citet{2007MNRAS.376.1251G}. Each field was divided into 61 smaller facets for the 323\,MHz images and into 31 facets for the 608\,MHz images. The facets were imaged separately, each with a different assumed phase centre, and then recombined into a hexagonal grid. In each case the total area covered by the facets is larger than the full width at half-maximum (FWHM) of the GMRT primary beam which allows bright sources well outside of the observed region to be cleaned from the images. This technique was used to better clean the full GMRT field of view in order to create the catalogue of sources. The task \textsc{setfc} was used to create a list of facet positions for use in \textsc{imagr}. Images were made originally with a pixel size of 3\,arcsec at 323\,MHz and 1.5\,arcsec at 608\,MHz to adequately sample the native synthesised beams. 

Each field went through three iterations of phase self-calibration using a model dominated by the bright sources in the field at 10, 3 and 1\,min intervals, and then a final round of self-calibration correcting both phase and amplitude errors at 10\,min intervals \citep[following][]{2007MNRAS.376.1251G}. The overall amplitude gain was held constant so that the flux density of sources was unaffected. These self-calibration steps improved the noise levels up to 30\,per~cent and significantly reduced the residual side lobes around bright sources. 

Final images of each field were originally produced with \textsc{robust} set to 0 within \textsc{imagr} to optimise the trade-off between angular resolution and sensitivity \citep[see results presented in][]{2016MNRAS.459.1248A}. To create a catalogue of sources within the entire field, the data were re-imaged after the final amplitude and phase self-calibration step for each field at each frequency in order to obtain matching beams and cell sizes for simultaneous source fitting. Specifically, all final images were made using a cell size of 1.5\,arcsec and the dimensions of the \textsc{clean} restoring beams of the original 323\,MHz images presented in \citet[][re-listed in Table~\ref{tab:srclist} of this paper]{2016MNRAS.459.1248A} for each respective field at both 323 and 608\,MHz. 

\begin{figure*}
\subfloat[L1551~IRS~5 field at 323\,MHz]{\includegraphics[width=0.39\textwidth]{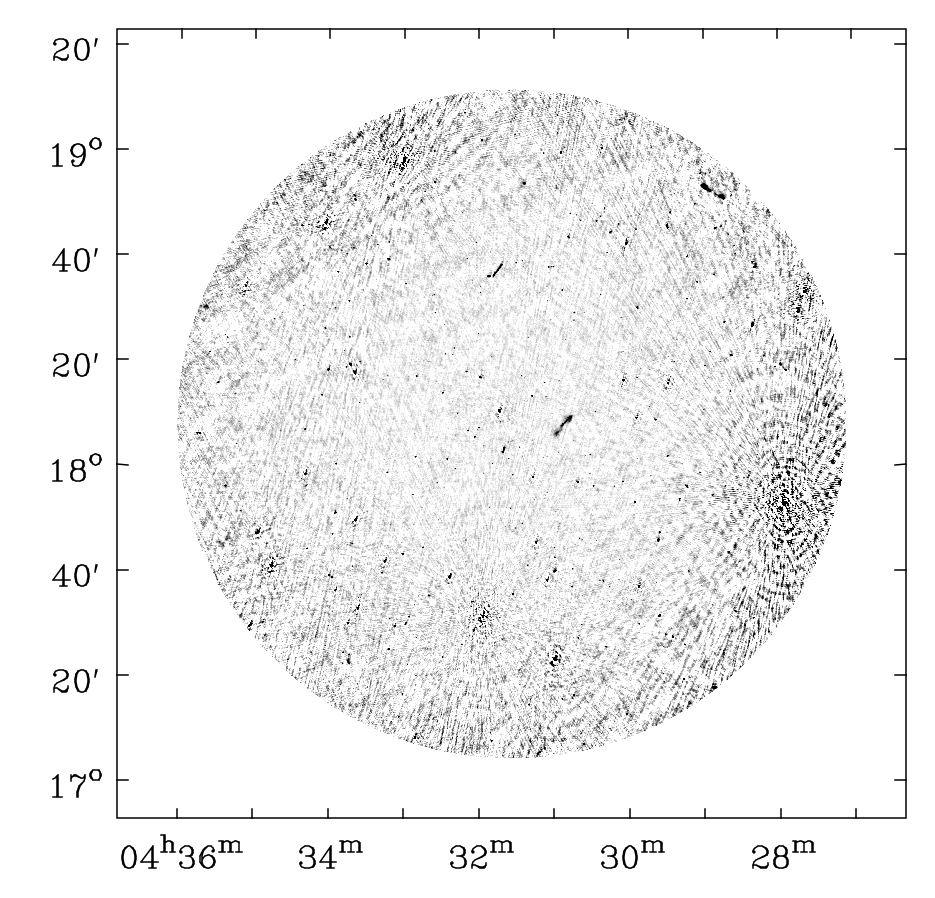}\label{fig:L1551_323MHz_PB}}
\subfloat[L1551~IRS~5 field at 608\,MHz]{\includegraphics[width=0.38\textwidth]{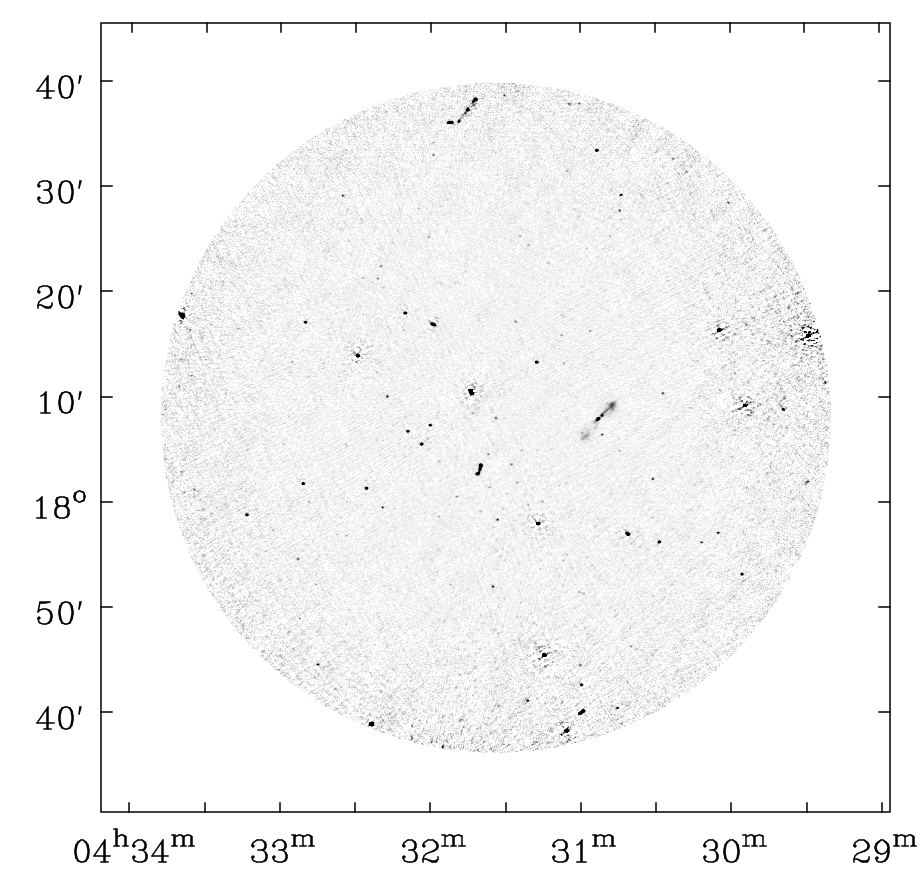}} \\
\subfloat[T~Tau field  at 323\,MHz]{\includegraphics[width=0.39\textwidth]{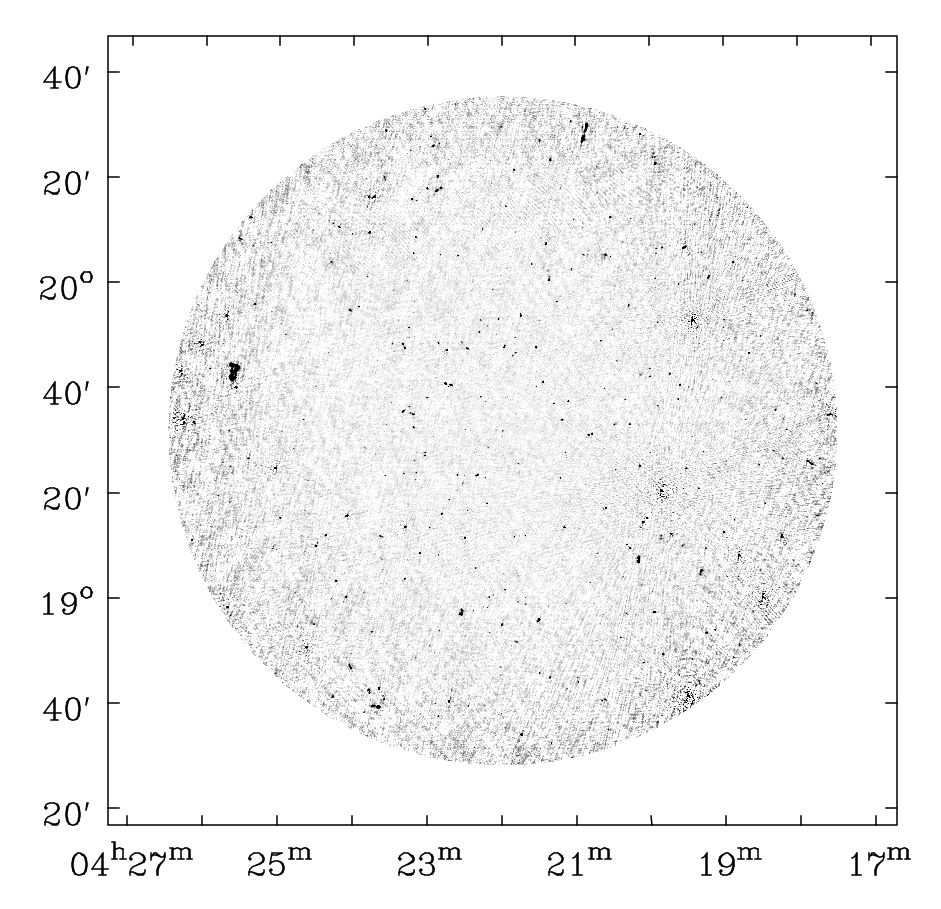}}
\subfloat[T~Tau field at 608\,MHz]{\includegraphics[width=0.39\textwidth]{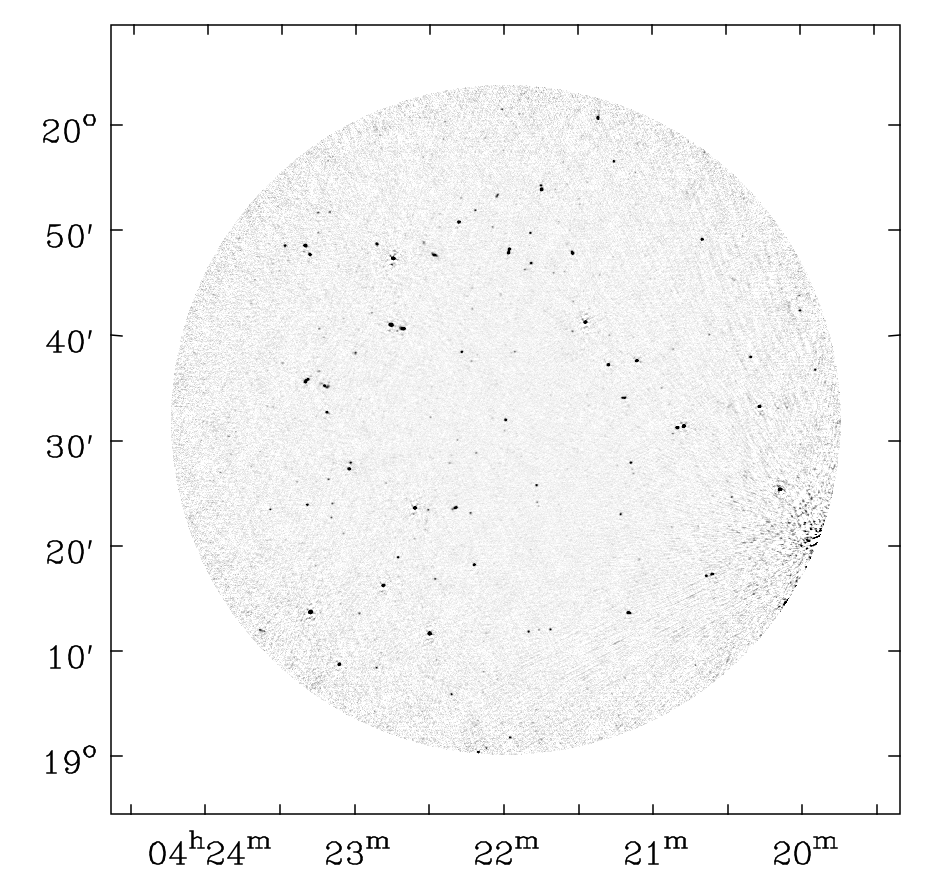}} \\
\subfloat[DG~Tau field  at 323\,MHz]{\includegraphics[width=0.39\textwidth]{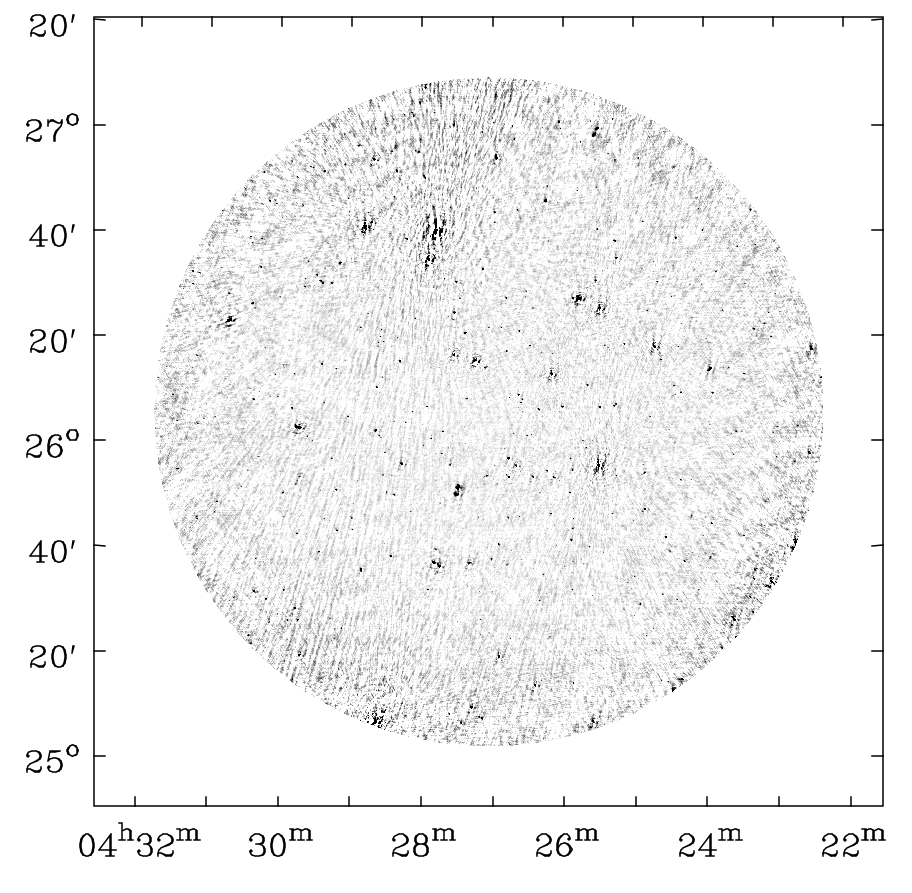}}
\subfloat[DG~Tau field at 608\,MHz]{\includegraphics[width=0.39\textwidth]{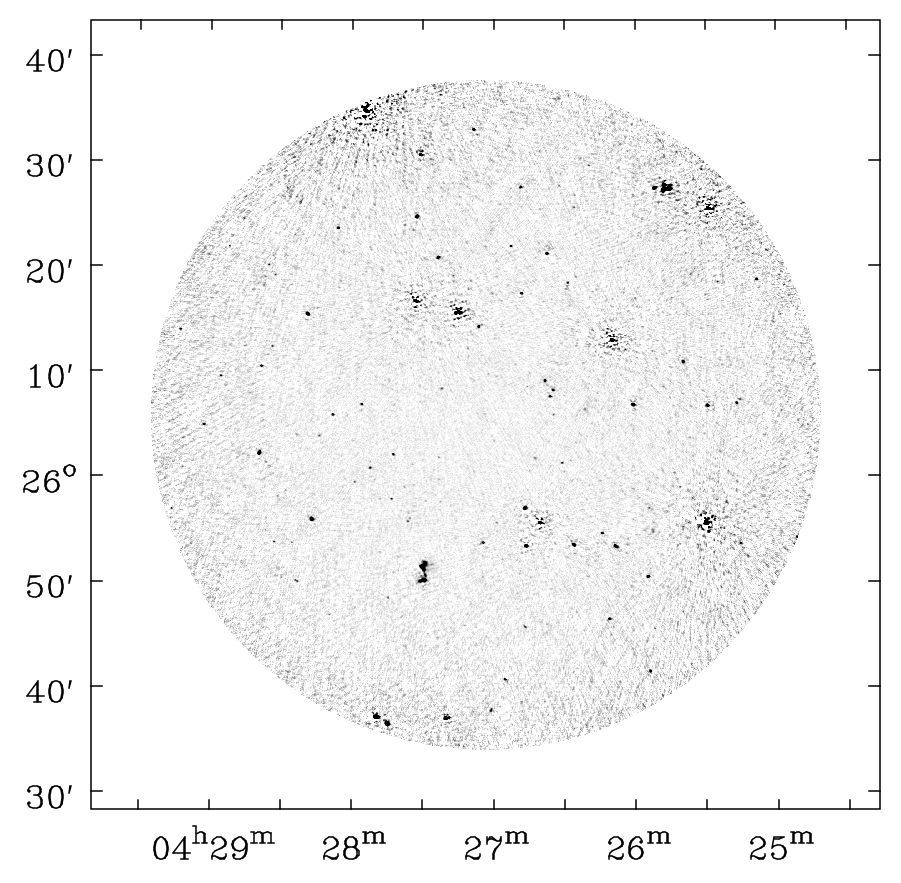}}
\caption{The full, \textsc{flatn}, primary-beam-corrected images of each target field at 323\,MHz (left column) and 608\,MHz (right column). The GMRT primary beam has a FWHM of approximately 85\,arcmin at 323\,MHz and 44\,arcmin at 608\,MHz and the maps are cut-off at the point where the primary beam correction factor dropped to 20\,per~cent of its central value. Greyscale ranges from $-0.1$ to 2\,mJy and residual calibration artefacts can be seen near the very bright sources, see e.g. the southwest region in (a). Axes are J2000.0 coordinates.}
\label{fig:pbcor}
\end{figure*}

The resulting facets were then stitched together with \textsc{flatn} and corrected for the primary beam using an eighth-order polynomial with coefficients taken from the GMRT Observer's Manual\footnote[1]{http://www.ncra.tifr.res.in/ncra/gmrt/gmrt-users/observing-help-for-gmrt-users/gmrt-observers-manual-07-july-2015/view} within \textsc{pbcor} and cut-off at the point where the primary beam correction factor dropped to 20\,per~cent of its central value \citep[following][]{2007MNRAS.376.1251G}. The primary beam of the GMRT has a FWHM of approximately 85\,arcmin at 325\,MHz and 44\,arcmin at 610\,MHz. The full, \textsc{flatn}, primary-beam-corrected images for each field at 323 and 608\,MHz are presented in Fig.~\ref{fig:pbcor}.

We recovered root-mean-squared (rms) noise ($\sigma_{\rm rms}$) levels of order $\sim100\,\umu$Jy\,beam$^{-1}$ at 323\,MHz and $\sim50\,\umu$Jy\,beam$^{-1}$ at 608\,MHz within a local patch of sky at the centre of each pointing after correction for the GMRT primary beam using the AIPS task {\sc imean} (see Table~\ref{tab:srclist}). The theoretical noise limit (in Jy\,beam$^{-1}$) for the observations is given by
\begin{equation}
\sigma_{\rm rms,th} = \frac{T_{\rm sys}}{G\,\sqrt{n_{b}\,n_{\rm pol}\,\Delta\nu\,\tau}} \times f,
\end{equation}
where $T_{\rm sys}$ is the system temperature (in K), $G$ is the antenna gain (in K\,Jy$^{-1}$), $n_{b}$ is the number of baselines ($n_{b}=n_{a}(n_{a}-1)/2$, where $n_{a}$ is the number of antennas), $n_{\rm pol}$ is the number of polarisations (2 in this case), $\Delta\nu$ is the total bandwidth of the observation (in Hz), $\tau$ is the observing time (in s) and $f$ is the ``fudge factor'' suggested by the GMRT sensitivity calculator (factors of $\sqrt{5}$ and $\sqrt{2}$ for 325 and 610 MHz observing modes, respectively). The GMRT Observer's Manual gives the system temperatures to be 106 and 102\,K for the 325 and 610\,MHz observing modes, respectively, while the antenna gains are 0.32\,K\,Jy$^{-1}$\,Antenna$^{-1}$. Using the observation times listed in Table~\ref{tab:srclist}, the expected sensitivities of the observations are $47\,\umu$Jy\,beam$^{-1}$ for the T~Tau field at 325\,MHz, $35\,\umu$Jy\,beam$^{-1}$ for L1551 IRS 5 and DG~Tau at 325\,MHz (due to different on-source observing time), and $35\,\umu$Jy\,beam$^{-1}$ for all fields at 610\,MHz. The achieved $\sigma_{\rm rms}$ values listed in Table~\ref{tab:srclist} are in relatively good agreement with these theoretical values for the 608\,MHz data, while the 323\,MHz sensitivities are a factor of 2-3 higher than anticipated. This is likely caused by the numerous extended bright sources with residual calibration artefacts visible in Fig.~\ref{fig:pbcor}.

As described in Section~2 of \citet{2016MNRAS.459.1248A}, a small systematic offset may be present between the 323 and 608\,MHz maps of the L1551~IRS~5 and DG~Tau fields. A number of bright, compact (presumably extragalactic) sources in each field were fitted using \textsc{jmfit} in \textsc{AIPS} at both 323 and 608\,MHz to compare the absolute positions. The positions at 323\,MHz were found to differ from those at 608\,MHz by approximately 0.86\,arcsec in declination for the L1551~IRS~5 field and 2\,arcsec in declination for the DG~Tau field. Positions are tied to the assumed position of the phase calibrator, so the likely cause for an offset in declination is a different line of sight through the ionosphere (which is up to $\approx6^\circ$ in separation for DG~Tau). This effect should be worse at 323\,MHz than 608\,MHz. We therefore apply the same shifts as in \citet{2016MNRAS.459.1248A} to the re-imaged L1551~IRS~5 and DG~Tau fields at 323\,MHz to make the source positions consistent with those at 608\,MHz.

\vspace{-10pt}
\section{Source Catalogue Creation}
\label{sec:cat_creation}

Source catalogues were created from each final image using the Python Blob Detection and Source Measurement (\textsc{PyBDSM}) package \citep{2015ascl.soft02007M}. Originally developed for use on LOFAR data, \textsc{PyBDSM} is designed to decompose low frequency radio images into source catalogues by mapping the variation of noise levels across the image and fitting significant `islands' of flux with appropriate Gaussians.

\subsection{Source Fitting}
\label{sec:sf}

Island detection was performed within \textsc{PyBDSM} using small sub-images to calculate the local rms deviation. The size of the sub-image was automatically adapted to account for the increased artifacting that may occur close to bright sources. Any islands with a peak of above $5\,\sigma_{\rm rms}$ were passed to the fitting routine with the extent of the island clipped at $3\,\sigma_{\rm rms}$. While the size of the local sub-images varied across individual fields, as well as across the three fields considered, the island thresholds of 5 and $3\,\sigma_{\rm rms}$ were kept constant across the survey. 

\textsc{PyBDSM} was used to model the emission from each significant island with a series of Gaussians. Extended sources were fitted with multiple Gaussians. These fits were then exported as source catalogues for further processing to create the final data products, including calculating the spectral index for sources detected at both frequencies. The peak and integrated flux densities were read from the \textsc{Peak\_flux} and \textsc{Total\_flux} columns of the \textsc{PyBDSM} catalogue. Errors on the flux densities are manually calculated as $\sigma_{S_{\nu}} = \sqrt{(0.05S_{\nu})^2+\sigma_{\rm{fit}}^2}$, where $0.05S_{\nu}$ is an estimated 5\,per~cent absolute calibration error on the flux density $S_{\nu}$ and $\sigma_{\rm fit}$ is the fitting error returned from \textsc{PyBDSM} via the columns \textsc{E\_Peak\_flux} and \textsc{E\_Total\_flux}. The individual frequency catalogues of each field were then combined and cross-referenced with the NVSS \citep{1998AJ....115.1693C} at 1.4\,GHz. 

\begin{figure*}
\centering
\subfloat[GMRT-TAU J043051.62+180819.1 and GMRT-TAU J043053.60+180752.4]{\includegraphics[width=0.34\textwidth]{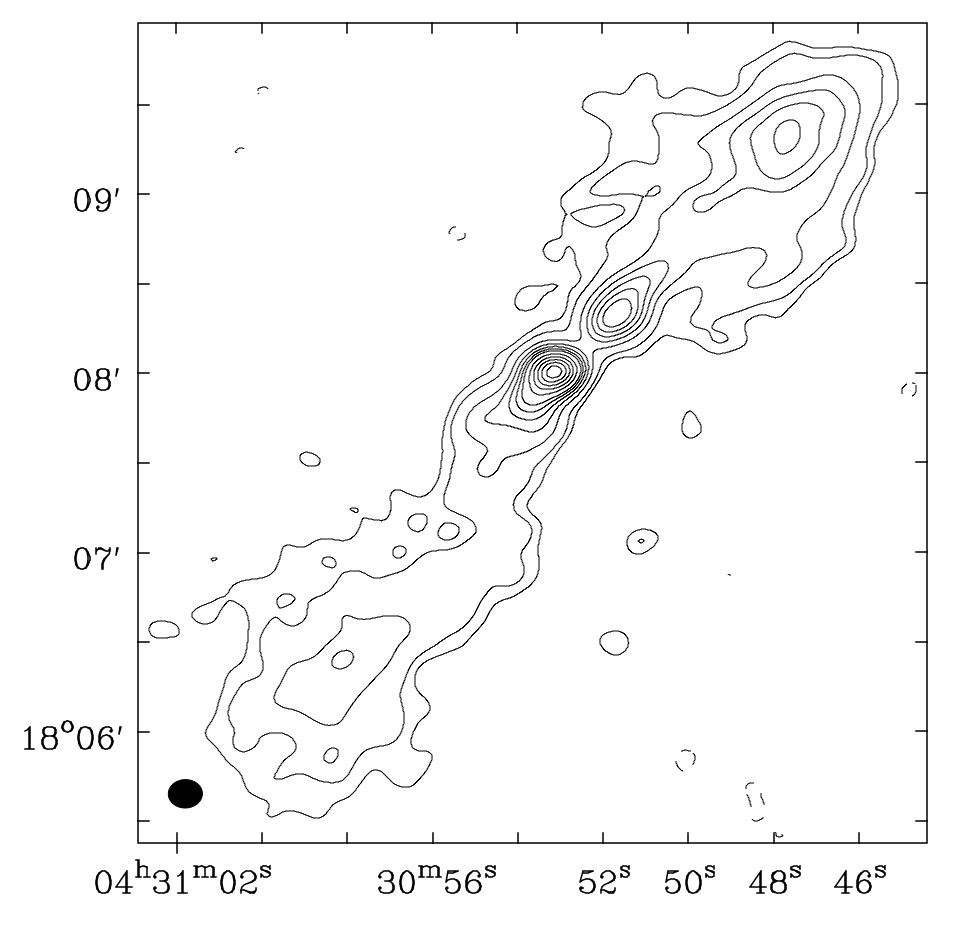}} \qquad
\subfloat[GMRT-TAU J043140.25+180325.8 and GMRT-TAU J043141.23+180248.3]{\includegraphics[width=0.32\textwidth]{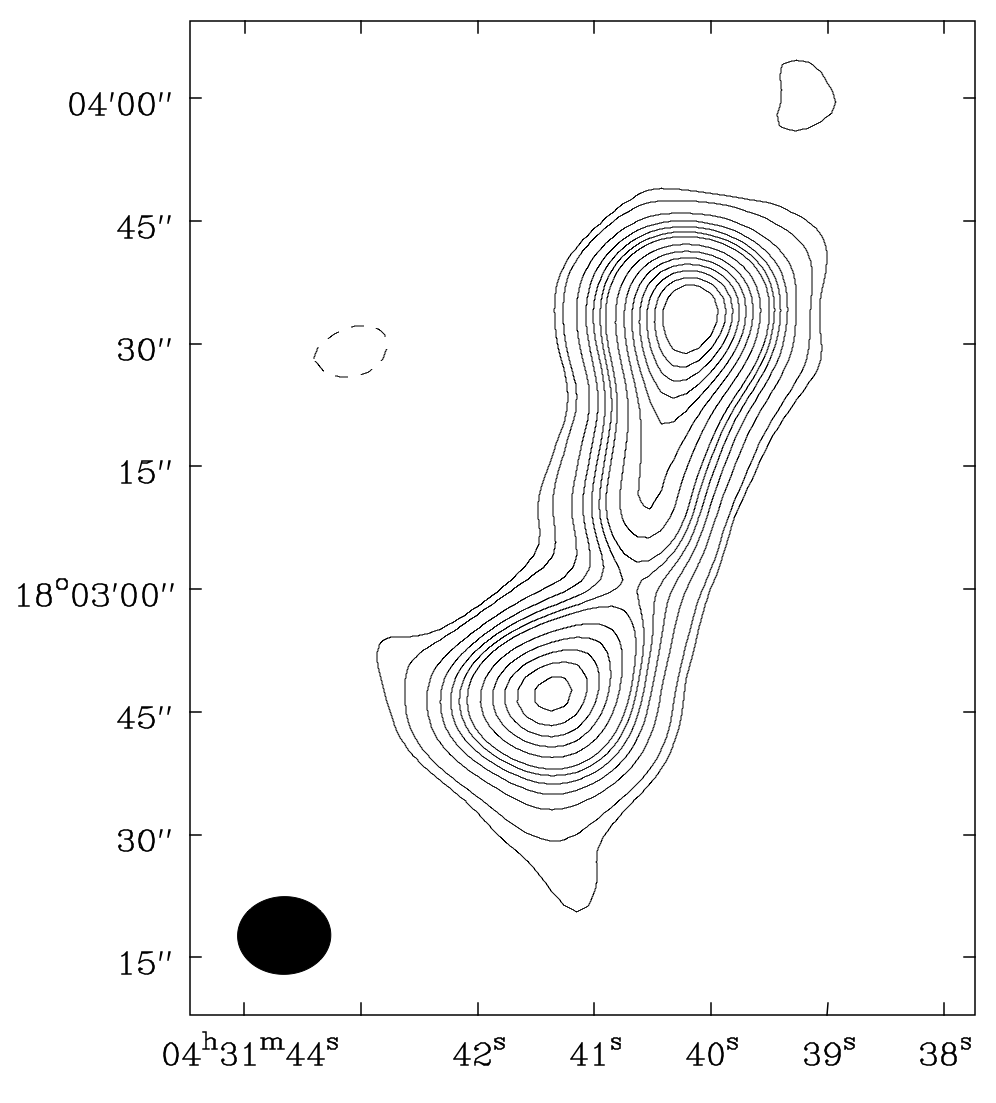}} \\
\subfloat[GMRT-TAU J043142.45+183818.0, GMRT-TAU J043145.33+183722.0, GMRT-TAU J043148.86+183617.1 and GMRT-TAU J043151.93+183609.7]{\includegraphics[width=0.34\textwidth]{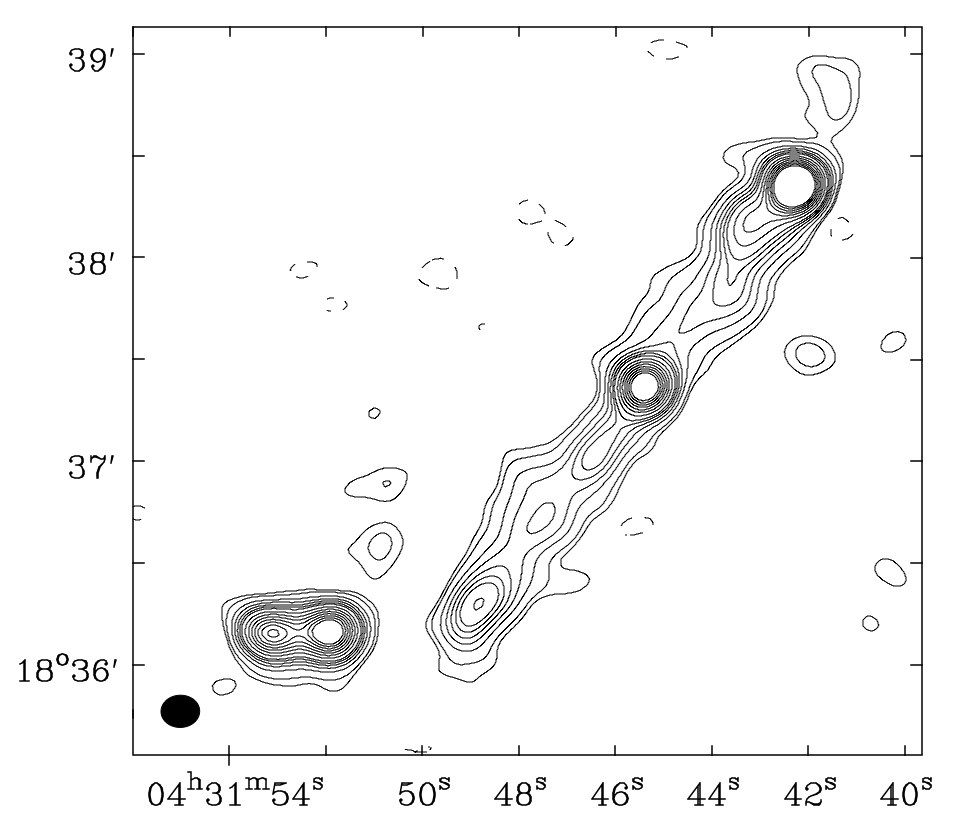}} \qquad
\subfloat[GMRT-TAU J042240.95+194046.5, GMRT-TAU J042243.28+194033.4 and GMRT-TAU J042245.78+194108.5]{\includegraphics[width=0.36\textwidth]{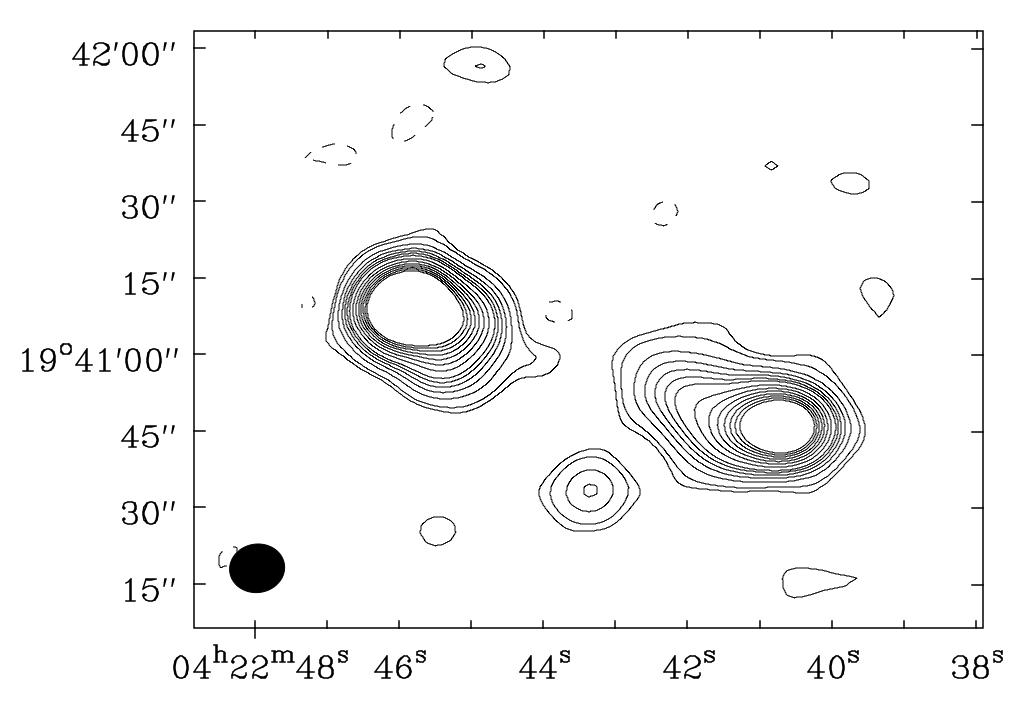}} \\
\subfloat[GMRT-TAU J042232.37+185760.0 and GMRT-TAU J042233.00+185717.3]{\includegraphics[width=0.34\textwidth]{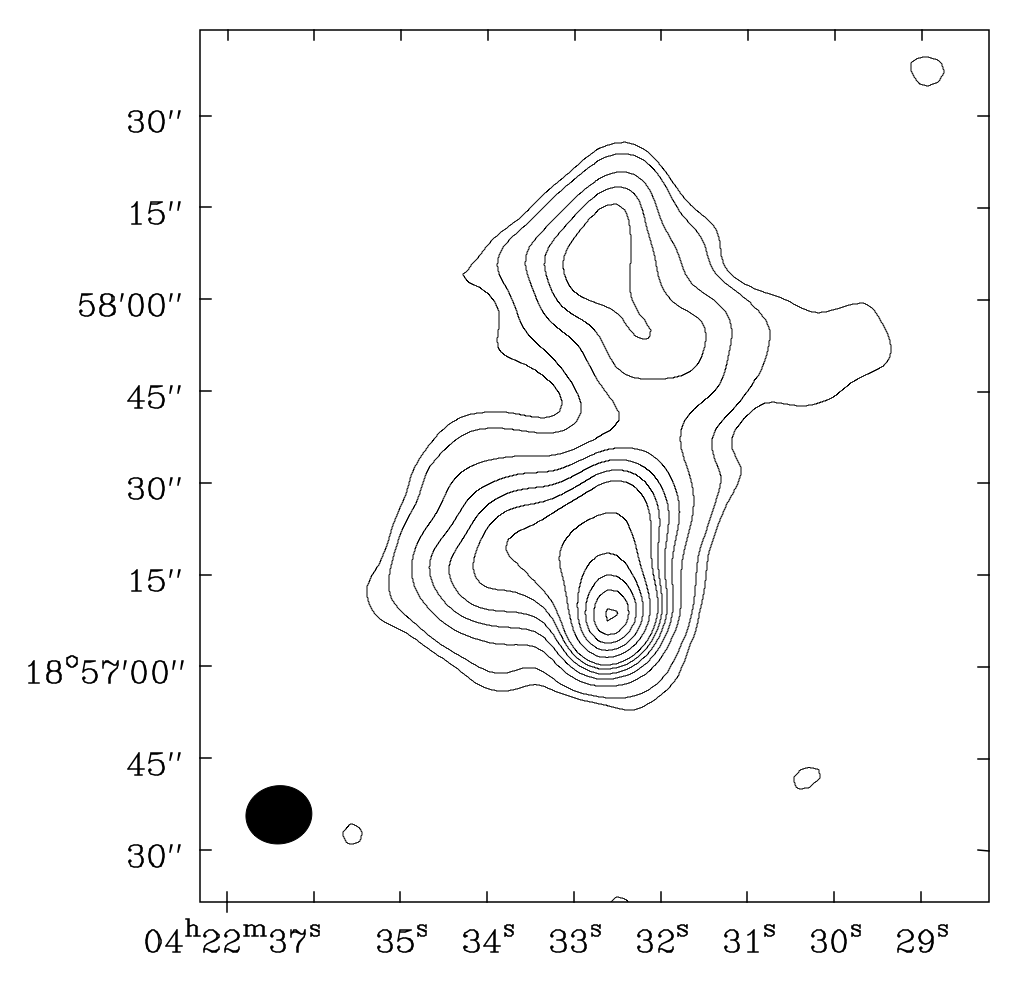}} \qquad
\subfloat[GMRT-TAU J042729.21+255046.0, GMRT-TAU J042730.16+255121.0 and GMRT-TAU J042730.32+255013.2]{\includegraphics[width=0.32\textwidth]{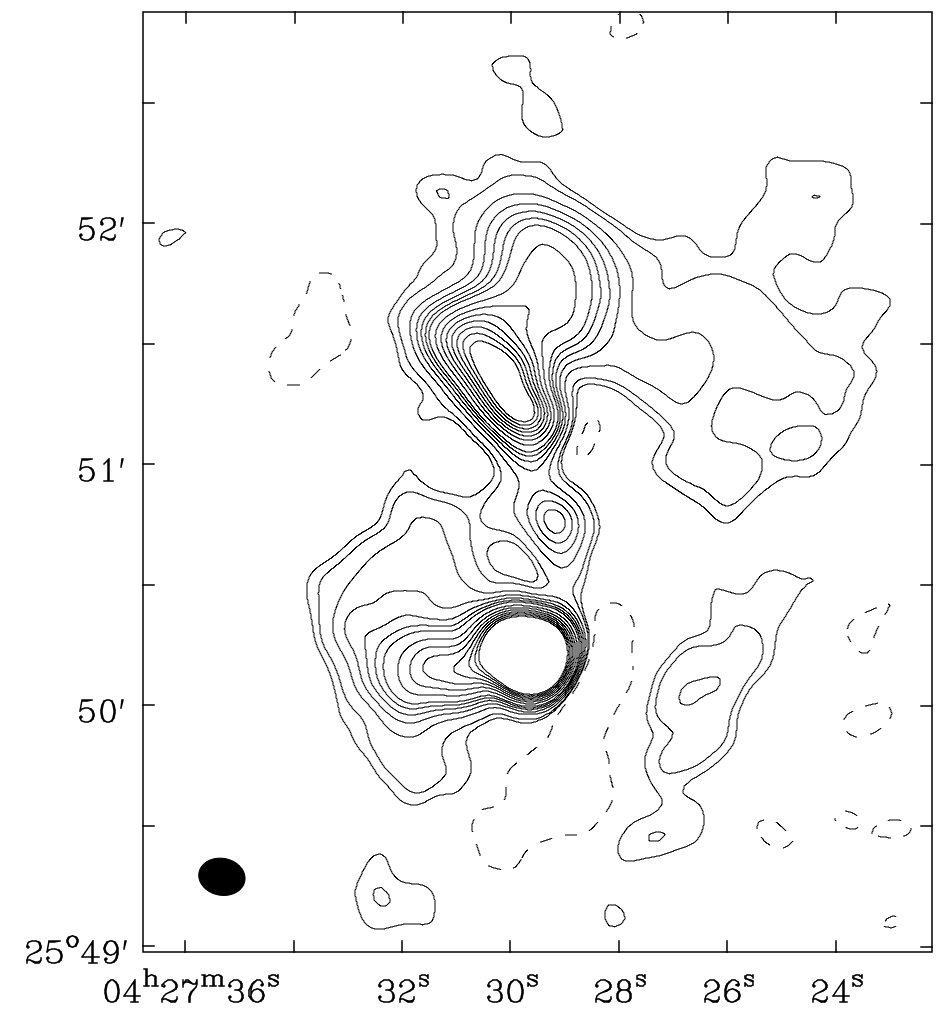}}
\caption{A sample of extended objects in the GMRT-TAU fields at 323\,MHz. (a), (b) and (c) are located within the L1551~IRS~5 field, (d) and (e) are located within the T~Tau field and (f) is located within the DG~Tau field. Contours are plotted as $-3$, 3, 5, 10, 15, 20, 25, 30, 40, 50, 60, 70, 80, 90 and $100\times\sigma_{\rm rms}$. The values for the local $\sigma_{\rm rms}$ are (a) 158, (b) 148, (c) 188, (d) 124, (e) 137 and (f) $142\,\umu$Jy\,beam$^{-1}$ at 323\,MHz. Axes are J2000.0 coordinates and the \textsc{clean} restoring beam is shown as a filled ellipse in the bottom left corner of each image (see Table~\ref{tab:srclist} for dimensions).}
\label{fig:maps}
\end{figure*}

Fig.~\ref{fig:pbcor} gives an indication of the quality of each of the three fields and the distributions of source sizes and brightnesses within them. Most sources detected are unresolved at both 323 and 608\,MHz (as the lower resolution beam was used for source fitting), although there are some objects present with extended structures. We showcase a sample of extended radio galaxies detected at 323\,MHz in Fig.~\ref{fig:maps}.

\subsection{Spectral Indices}
\label{sec:spin}

Sources detected at both GMRT frequencies within a proximity of less than 5\,arcsec \citep[chosen based on the native resolution of the 608\,MHz data, see][]{2016MNRAS.459.1248A} were considered to be the same source and fitted with a spectral index $\alpha_{\rm GMRT}$ using
\begin{equation}
\alpha_{\rm GMRT}=\frac{\log(S_{323}/S_{608})}{\log({323/608)}}
\label{eqn:sp_indx}
\end{equation}
where $S_{323}$ and $S_{608}$ are the flux densities at frequencies 323 and 608\,MHz, respectively. These spectral indices were calculated using the fitted integrated fluxes for resolved sources and peak fluxes for unresolved sources, where a source with a major axis of greater than 10\,arcsec was considered resolved. The error on the spectral index, $\sigma_{\alpha_{\rm GMRT}}$, was calculated as
\begin{equation}
\sigma_{\alpha_{\rm GMRT}}=\frac{1}{|\log(323/608)|} \left[ \left( \frac{\sigma_{S_{323}}}{S_{323}}\right)^{2} + \left(\frac{\sigma_{S_{608}}}{S_{608}} \right)^{2} \right]^{1/2}
\end{equation}
where $\sigma_{S_{323}}$ and $\sigma_{S_{608}}$ are the errors on the flux densities at 323 and 608\,MHz, respectively. 

An NVSS detection within 5\,arcsec of a GMRT detection was considered a match and a spectral index $\alpha_{\rm NVSS}$ between the two frequencies was calculated following Equation~\ref{eqn:sp_indx} (using the NVSS flux density at 1.4\,GHz for the higher flux and frequency). In cases where a source was detected at both GMRT frequencies and in the NVSS catalogue, a linear fit was used to calculate the spectral index $\alpha_{\rm NVSS}$ using all three fluxes.

\subsection{Final Catalogue}
\label{sec:fc}

An average of 605 sources were detected within each individual field at 323\,MHz and an average of 229 sources were detected within each individual field at 608\,MHz. There is not a significant variance between the three fields at 608\,MHz, however there are of order 100 less sources detected within the L1551~IRS~5 field at 323\,MHz compared to the T~Tau and DG~Tau fields. This is likely to be due to the increased noise levels caused by very bright sources (see e.g. the southwest region in Fig.~\ref{fig:L1551_323MHz_PB}). The breakdown of source counts from each individual field is presented in Table~\ref{table:source_counts}.

\begin{table}
\centering
\caption[Source counts.]{Source counts for the full catalogue of the three target regions in Taurus at 323 and 608\,MHz. This corresponds to a total of 2062 unique sources.}
\label{table:source_counts}
\begin{tabular}{lccc}
\hline
Field & 323\,MHz & 608\,MHz & Matches \\
\hline
L1551~IRS~5 & 507 & 224 & 136 \\
T~Tau & 689 & 261 & 175 \\
DG~Tau & 619 & 202 & 129 \\
\hline
Total & 1815 & 687 & 440 \\
\hline
\end{tabular}
\end{table}

In total, 1815 sources were detected across the three fields combined at 323\,MHz and 687 sources were detected at 608\,MHz. \citet{2007MNRAS.376.1251G} detected 3944 sources with the GMRT over 4 square degrees at 610\,MHz. Correcting these authors' source counts for the difference in sensitivity using their published analysis of source count variation with peak brightness, and the difference in observing area, we expect about 700 source detections within the total observing area of this survey at 608\,MHz. This is consistent with our 687 detections across the combined fields at 608\,MHz.

A total of 440 sources detected across the fields were matched between the two frequencies, yielding a total unique source count of 2062 detections which make up the final catalogue. A sample of the final catalogue (containing 50 sources within the T~Tau field) is presented in Table~\ref{table:cat} and the full catalogue can be found in the Supplementary Material through the online version of this article. The following are brief descriptions of each column:

Column~(1): The IAU designation of the source in the form of GMRT-TAU J\textit{hhmmss.ss}+\textit{ddmmss.s} where J represents J2000.0 coordinates, \textit{hhmmss.ss} represents Right Ascension (RA) in hours, minutes and truncated hundredths of seconds, and +\textit{ddmmss.s} represents Declination (Dec) in degrees, arcminutes and truncated tenths of arcseconds. 

Columns~(2)--(3): The source position RA and Dec as calculated from the position of the peak of the fitted Gaussian or, in the case of a multi-Gaussian source, the centroid of the combined source. In the case where a source was detected at both 323 and 608\,MHz, the high frequency position is reported as the 608\,MHz positions were found to be more consistent with the NVSS (discussed further in a later Section of this paper).

Columns~(4)--(7): Information relating to a source detection at 323\,MHz. Columns~(4)--(5) give peak and integrated flux densities measured by \textsc{PyBDSM} (\textsc{Peak\_flux} and \textsc{Total\_flux}, respectively) where all errors quoted are $1\,\sigma_{S_\nu}$ (see error calculation in Section~\ref{sec:sf}). Column~(6) gives the \textsc{PyBDSM} parameter output for \textsc{Resid\_Isl\_rms}, which is the average residual background rms value of the island (i.e. the local island noise). Column~(7) lists the \textsc{PyBDSM} parameter output for \textsc{S\_Code} which is a code that defines the multiplicity of the source structure in terms of Gaussian components. An \textsc{S\_Code} of `S' corresponds to a single-Gaussian source that is the only source in the island of emission, `C' corresponds to a single-Gaussian source in an island with other sources and `M' corresponds to a multi-Gaussian source. 

Columns~(8)--(11): Information relating to a source detection at 608\,MHz. Columns~(8)--(9) give peak and integrated flux densities, Column~(10) gives the \textsc{PyBDSM} parameter output for \textsc{Resid\_Isl\_rms} and Column~(11) lists the \textsc{PyBDSM} parameter output for \textsc{S\_Code}. 

Column~(12): The spectral index between the two GMRT frequencies ($\alpha_{\rm GMRT}$) for sources detected at both 323 and 608\,MHz. 

Columns~(13)--(14):  Cross-reference with the NVSS. Column~(13) lists if a source detected in the catalogue has a corresponding NVSS detection, where `L' indicates a match with the lower frequency (323\,MHz) GMRT data only, `H' indicates a match with the higher frequency (608\,MHz) GMRT data only and `B' indicates a match with both GMRT frequencies. The corresponding GMRT to NVSS spectral index $\alpha_{\rm NVSS}$ listed in Column~(14) is generated using two frequencies in the case of a `L' or `H' match, and three frequencies in the case of a `B' match. See Section~\ref{sec:spin} above for details on the spectral index calculations. 

Column~(15): The known counterparts of a source cross-referenced with GBS-VLA \citep[][denoted by a G]{2015ApJ...801...91D}, NVSS \citep[][denoted by an N]{1998AJ....115.1693C}, the XMM-Newton extended survey of the Taurus molecular cloud \citep[XEST,][denoted by an X]{2007A&A...468..353G}, the Two Micron All Sky Survey \citep[2MASS,][denoted by 2M]{2006AJ....131.1163S} and the \textit{Spitzer} c2d Legacy Program \citep[][denoted by an S]{2003PASP..115..965E}. 

It should be noted that \textsc{PyBDSM} reports a number of additional parameters which are not part of the catalogue but which are made available in a machine-readable format as part of the released data products (see Section~\ref{sec:dp}), along with the scripts used to make the final catalogue and validation plots. A full description of the parameters fitted by \textsc{PyBDSM} is available in its online documentation\footnote[2]{http://www.astron.nl/citt/pybdsm/}.

\begin{landscape}
\begin{table}
\centering
\caption[Catalogue sample.]{A sample of 50 entries from the final GMRT-TAU catalogue, the full table is made available in the Supplementary Material through the online version of this article. Peak and integrated fluxes, local noises and \textsc{PyBDSM} codes are reported separately for each frequency and are omitted if no detection is made. Where detections at multiple frequencies occur the corresponding spectral index is reported. See Section~\ref{sec:fc} for a detailed description of each column. This table is also in a machine-readable format on the project website with a corresponding ReadMe file to describe the slightly differing columns, see Section~\ref{sec:dp}.}
\label{table:cat}
\tiny
\begin{tabular}{cccccccccccccccc}
\hline
 & & & \multicolumn{4}{c}{323\,MHz} & & \multicolumn{4}{c}{608\,MHz} &  &  &  & \\ \cmidrule{4-7}\cmidrule{9-12}
Name & $\alpha_{\rm J2000.0}$ & $\delta_{\rm J2000.0}$ & $S_{\rm peak}$ & $S_{\rm int}$ & \textsc{Resid\_Isl\_rms} & \textsc{S\_Code} & & $S_{\rm peak}$ & $S_{\rm int}$ & \textsc{Resid\_Isl\_rms} & \textsc{S\_Code} & $\alpha_{\rm GMRT}$ & NVSS & $\alpha_{\rm NVSS}$ & Known \\
GMRT-TAU & ($^{\rm h}:^{\rm m}:^{\rm s}$) & ($^{\circ}$:$'$:$''$) & (mJy\,beam$^{-1}$) & (mJy) & ($\umu$Jy\,beam$^{-1}$) &  & & (mJy\,beam$^{-1}$) & (mJy) & ($\umu$Jy\,beam$^{-1}$) &  &  &  &  & counterparts \\
\hline
J042200.00+185517.2&04:22:00.001&+18:55:17.18&$26.21\pm1.33$&$42.22\pm2.15$&$108$&M&&-&-&-&-&-&-&-&-\\
J042200.28+195743.1&04:22:00.283&+19:57:43.06&$0.98\pm0.13$&$1.14\pm0.21$&$11$&S&&$0.64\pm0.10$&$0.41\pm0.18$&$23$&S&$-0.92\pm0.85$&-&-&-\\
J042200.53+183822.9&04:22:00.526&+18:38:22.94&$1.47\pm0.28$&$2.74\pm0.42$&$55$&S&&-&-&-&-&-&-&-&-\\
J042200.92+184709.3&04:22:00.918&+18:47:09.30&$1.93\pm0.22$&$2.81\pm0.34$&$43$&C&&-&-&-&-&-&-&-&-\\
J042200.93+200137.4&04:22:00.931&+20:01:37.40&$2.06\pm0.17$&$2.16\pm0.26$&$30$&S&&$1.84\pm0.18$&$1.76\pm0.28$&$93$&S&$-0.32\pm0.71$&-&-&-\\
J042200.96+202954.2&04:22:00.964&+20:29:54.25&$3.47\pm0.39$&$7.54\pm0.62$&$106$&S&&-&-&-&-&-&L&$-0.81\pm0.31$&N \\
J042201.08+184541.2&04:22:01.079&+18:45:41.17&$1.76\pm0.23$&$2.88\pm0.35$&$52$&S&&-&-&-&-&-&-&-&-\\
J042201.49+194404.2&04:22:01.494&+19:44:04.16&-&-&-&-&&$0.43\pm0.06$&$0.34\pm0.11$&$19$&C&-&-&-&-\\
J042201.54+194354.5&04:22:01.536&+19:43:54.48&-&-&-&-&&$0.30\pm0.06$&$0.30\pm0.10$&$19$&C&-&-&-&-\\
J042201.69+184717.7&04:22:01.688&+18:47:17.73&$1.54\pm0.21$&$2.32\pm0.32$&$43$&C&&-&-&-&-&-&-&-&-\\
J042202.16+192800.3&04:22:02.162&+19:28:00.31&$0.53\pm0.11$&$0.54\pm0.19$&$5$&S&&-&-&-&-&-&-&-&-\\
J042202.82+190728.3&04:22:02.820&+19:07:28.27&$0.92\pm0.14$&$1.74\pm0.21$&$35$&S&&-&-&-&-&-&-&-&2M\\
J042202.88+195325.7&04:22:02.880&+19:53:25.75&$3.52\pm0.21$&$8.50\pm0.50$&$46$&M&&$1.88\pm0.13$&$4.33\pm0.25$&$117$&S&-&B&$-0.81\pm0.24$&N  2M\\
J042202.95+193530.7&04:22:02.948&+19:35:30.66&-&-&-&-&&$0.28\pm0.06$&$0.33\pm0.09$&$20$&S&-&-&-&-\\
J042202.98+191628.0&04:22:02.981&+19:16:27.99&-&-&-&-&&$0.28\pm0.07$&$0.41\pm0.11$&$24$&S&-&-&-&-\\
J042203.97+201134.3&04:22:03.971&+20:11:34.27&$0.96\pm0.16$&$1.14\pm0.26$&$12$&S&&-&-&-&-&-&-&-&-\\
J042204.77+195025.8&04:22:04.774&+19:50:25.85&$1.56\pm0.14$&$1.66\pm0.21$&$22$&S&&$0.86\pm0.09$&$1.07\pm0.13$&$63$&S&$-0.70\pm0.63$&-&-&-\\
J042204.90+193216.3&04:22:04.902&+19:32:16.29&$1.31\pm0.13$&$2.57\pm0.20$&$41$&S&&$0.41\pm0.06$&$0.51\pm0.09$&$21$&S&$-2.54\pm0.67$&-&-&-\\
J042206.37+201308.5&04:22:06.365&+20:13:08.54&$1.16\pm0.20$&$1.82\pm0.30$&$38$&S&&-&-&-&-&-&-&-&-\\
J042207.24+190053.8&04:22:07.235&+19:00:53.82&$2.18\pm0.18$&$2.41\pm0.27$&$17$&S&&$1.63\pm0.20$&$1.41\pm0.34$&$114$&S&$-0.85\pm0.97$&-&-&-\\
J042207.32+191150.3&04:22:07.324&+19:11:50.28&-&-&-&-&&$0.48\pm0.09$&$0.69\pm0.13$&$21$&S&-&-&-&-\\
J042207.78+195854.0&04:22:07.781&+19:58:54.02&$1.67\pm0.16$&$3.50\pm0.26$&$67$&S&&$0.65\pm0.12$&$0.83\pm0.19$&$29$&S&$-2.27\pm0.87$&-&-&-\\
J042208.33+190809.2&04:22:08.331&+19:08:09.24&$1.35\pm0.15$&$1.44\pm0.23$&$30$&S&&$0.92\pm0.11$&$0.66\pm0.19$&$31$&S&$-0.71\pm0.71$&-&-&-\\
J042208.60+201913.2&04:22:08.595&+20:19:13.23&$1.02\pm0.20$&$1.66\pm0.30$&$27$&S&&-&-&-&-&-&-&-&-\\
J042209.14+195449.3&04:22:09.142&+19:54:49.28&$0.55\pm0.13$&$1.61\pm0.18$&$31$&S&&-&-&-&-&-&-&-&2M\\
J042209.39+202332.3&04:22:09.395&+20:23:32.34&$1.57\pm0.23$&$2.55\pm0.34$&$85$&S&&-&-&-&-&-&-&-&-\\
J042209.98+185346.9&04:22:09.979&+18:53:46.94&$2.06\pm0.19$&$2.75\pm0.29$&$36$&S&&-&-&-&-&-&-&-&-\\
J042210.44+190031.5&04:22:10.442&+19:00:31.51&$8.62\pm0.46$&$10.07\pm0.57$&$63$&S&&$5.26\pm0.33$&$5.45\pm0.44$&$186$&S&$-0.97\pm0.36$&-&-&-\\
J042210.44+190330.5&04:22:10.440&+19:03:30.49&$0.60\pm0.13$&$0.76\pm0.20$&$13$&S&&-&-&-&-&-&-&-&-\\
J042210.81+200623.7&04:22:10.810&+20:06:23.72&$0.78\pm0.17$&$0.85\pm0.27$&$10$&S&&-&-&-&-&-&-&-&-\\
J042210.96+193817.7&04:22:10.962&+19:38:17.71&$0.66\pm0.13$&$0.78\pm0.20$&$10$&S&&$0.36\pm0.06$&$0.32\pm0.10$&$16$&S&$-1.38\pm1.39$&-&-&-\\
J042211.35+192857.3&04:22:11.349&+19:28:57.32&$1.89\pm0.15$&$2.25\pm0.21$&$53$&C&&$1.01\pm0.07$&$0.97\pm0.10$&$44$&S&-&-&-&-\\
J042211.65+195734.3&04:22:11.646&+19:57:34.28&$0.87\pm0.13$&$0.99\pm0.20$&$11$&S&&-&-&-&-&-&-&-&-\\
J042211.66+185832.0&04:22:11.664&+18:58:32.04&$2.18\pm0.19$&$2.48\pm0.29$&$33$&S&&-&-&-&-&-&-&-&-\\
J042211.67+192914.6&04:22:11.665&+19:29:14.58&$0.39\pm0.12$&$0.78\pm0.17$&$53$&C&&-&-&-&-&-&-&-&-\\
J042211.74+195201.9&04:22:11.742&+19:52:01.91&$2.45\pm0.17$&$2.73\pm0.24$&$58$&S&&$2.09\pm0.13$&$1.94\pm0.17$&$59$&S&$-0.54\pm0.45$&-&-&-\\
J042212.12+191819.8&04:22:12.125&+19:18:19.81&$11.52\pm0.59$&$12.05\pm0.64$&$44$&S&&$8.96\pm0.46$&$8.91\pm0.47$&$77$&S&$-0.48\pm0.27$&B&$-0.48\pm0.01$&N \\
J042213.00+194239.0&04:22:12.996&+19:42:39.05&$1.32\pm0.12$&$1.16\pm0.20$&$10$&S&&$0.72\pm0.07$&$0.70\pm0.11$&$51$&S&$-0.79\pm0.82$&-&-&-\\
J042213.06+194438.0&04:22:13.058&+19:44:37.99&-&-&-&-&&$0.40\pm0.07$&$0.56\pm0.11$&$23$&S&-&-&-&-\\
J042213.25+193741.7&04:22:13.245&+19:37:41.73&$1.64\pm0.15$&$1.82\pm0.22$&$34$&S&&$0.80\pm0.07$&$0.77\pm0.10$&$30$&S&$-1.36\pm0.65$&-&-&-\\
J042213.42+190329.4&04:22:13.420&+19:03:29.42&$0.78\pm0.13$&$1.16\pm0.20$&$14$&S&&-&-&-&-&-&-&-&-\\
J042213.60+192314.7&04:22:13.602&+19:23:14.71&$2.58\pm0.18$&$3.24\pm0.32$&$23$&M&&$1.55\pm0.10$&$1.76\pm0.17$&$36$&M&-&-&-&-\\
J042215.11+190702.4&04:22:15.109&+19:07:02.39&$1.03\pm0.15$&$1.59\pm0.22$&$15$&S&&-&-&-&-&-&-&-&-\\
J042215.31+185838.0&04:22:15.310&+18:58:37.97&$1.63\pm0.19$&$1.92\pm0.29$&$29$&S&&-&-&-&-&-&-&-&-\\
J042215.89+201028.0&04:22:15.888&+20:10:27.96&$3.14\pm0.25$&$4.70\pm0.38$&$104$&S&&-&-&-&-&-&-&-&-\\
J042215.98+194138.2&04:22:15.980&+19:41:38.23&$0.53\pm0.10$&$0.44\pm0.18$&$3$&S&&-&-&-&-&-&-&-&-\\
J042217.20+193834.6&04:22:17.202&+19:38:34.58&$6.86\pm0.37$&$6.77\pm0.40$&$55$&S&&$4.48\pm0.24$&$4.42\pm0.25$&$80$&S&$-0.67\pm0.29$&B&$-0.71\pm0.05$&N  G\\
J042217.58+195312.0&04:22:17.585&+19:53:12.01&$2.78\pm0.18$&$4.45\pm0.34$&$46$&M&&$0.78\pm0.09$&$0.86\pm0.15$&$39$&S&-&-&-&-\\
J042218.38+195054.7&04:22:18.376&+19:50:54.72&$14.57\pm0.74$&$16.05\pm0.83$&$125$&S&&$11.60\pm0.59$&$11.96\pm0.62$&$104$&S&$-0.46\pm0.27$&B&$-0.41\pm0.03$&N \\
J042218.93+193011.7&04:22:18.933&+19:30:11.67&$1.18\pm0.12$&$1.26\pm0.18$&$27$&S&&$0.76\pm0.06$&$0.60\pm0.09$&$33$&S&$-0.81\pm0.58$&-&-&-\\
\hline
\end{tabular}
\end{table}
\end{landscape}

\vspace{-10pt}
\section{Discussion}
\label{sec:dis}

The results presented in \citet{2016MNRAS.459.1248A} focussed on the study of emission detected from the three target YSOs at the phase centres of each field (L1551~IRS~5, T~Tau and DG~Tau). In this work, the same set of observations has been used to make a catalogue of these regions towards the TMC at 323 and 608\,MHz. Although the TMC is not as densely populated as other star forming regions \citep[e.g. $\rho$~Ophiuchus, Orion;][]{2015ApJ...801...91D}, there are several other pre-main-sequence objects with known radio emission located within the observed fields. However, no additional protostellar objects were detected apart from the target YSOs.

\subsection{YSO detections}

We cross-referenced the GMRT-TAU catalogue with the GBS-VLA to search for additional known radio emitting YSOs within our fields. Catalogue sources GMRT-TAU~J043134.03+180803.9 and GMRT-TAU~J042159.51+193206.3 are identified as the target sources L1551~IRS~5 (GBS-VLA~J043134.16+180804.6) and T~Tau (GBS-VLA~J042159.43+193205.7), respectively. No additional known YSOs from the GBS-VLA were detected within the source finding criteria described in Section~\ref{sec:sf} (i.e. at the $5\,\sigma_{\rm rms}$ level). 

We also searched for sources within our catalogue which would have thermal spectral indices when compared with objects within the GBS-VLA. We find two thermal sources within the L1551~IRS~5 field other than L1551~IRS~5 itself: GMRT-TAU~J043150.37+182052.5 (GBS-VLA~J043150.44+182052.6) and GMRT-TAU~J043229.39+181359.8 (GBS-VLA~J043229.46+181400.2).

GMRT-TAU~J043229.39+181359.8 was classified as a Class~II transition disk candidate (Tau~L1551~13) by \citet{2009ApJS..184...18G} through application of mid-infrared colour-based methods to \textit{Spitzer} data. This object has also been detected with 2MASS, XEST, NVSS and the GBS-VLA. Based on its 4.5--7.5\,GHz properties, \citet{2015ApJ...801...91D} suggested this object (GBS-VLA~J043229.46+181400.2) is instead an extragalactic background source based on its low variability ($<50$\,per~cent), negative spectral index ($\alpha_{\rm 4.5\,GHz}^{\rm 7.5\,GHz}=-0.25\pm0.71$) and lack of proper motion. Its radio flux density is also large at centimetre wavelengths ($\sim60$\,mJy) compared with most other YSOs in the TMC ($\sim1$\,mJy). Combining the GMRT 323 and 608\,MHz data, NVSS data at 1.4\,GHz and the GBS-VLA data at 4.5 and 7.5\,GHz, we find a spectral index $\alpha_{\rm 323\,MHz}^{\rm 7.5\,GHz}=0.31\pm0.03$ for GMRT-TAU~J043229.39+181359.8. This spectral index is consistent with thermal, free--free radiation which is typically detected within star forming regions from YSO outflows. However, the relatively high flux density compared to other YSOs within the TMC and the lack of proper motion \citep[members exhibit proper motions of $\sim20$\,mas\,yr$^{-1}$, see e.g.][]{2009ApJ...698..242T} still suggests that this object is extragalactic and the thermal spectrum may be related to star formation activity. We follow \citet{2015ApJ...801...91D} and do not consider this object to be a Taurus YSO for the rest of this paper.

The only thermal source we detect within the T~Tau field when compared with the GBS-VLA is T~Tau itself. We detect two sources with apparent thermal spectral indices within the DG~Tau field: GMRT-TAU~J042452.49+264203.2 (GBS-VLA~J042452.48+264204.5) which has been classified as extragalactic by the Sloan Digital Sky Survey \citep[SDSS][]{2011yCat.2306....0A}, and GMRT-TAU~J042929.48+263151.6 (GBS-VLA~J042929.49+263152.8) which has been classified as a field star by \citet{2015ApJ...801...91D} but as a known galaxy by \citet{2011ApJS..196....4R} and SDSS. The GBS-VLA and SDSS properties of GMRT-TAU~J042929.48+263151.6 suggest that it is extragalactic. We therefore do not detect any YSOs in addition to the three target sources based on a cross-reference with the GBS-VLA. It is possible that targeted manual searches may reveal detections.

\subsection{Notable YSO non-detections}

Although no additional YSOs with known radio emission were detected within our fields, we emphasise that the work presented here is a by-product of a pathfinder project to detect the specific YSO target sources at the phase centre of each pointing, which was successful. We discuss some cases of notable YSO non-detections in the following paragraphs. 

Variability may partially account for the non-detections of additional YSOs within our single epoch observations. Non-thermal radio emission associated with active coronae (which we may expect to detect at such low frequencies) often exhibits high levels of variability \citep[>50\,per~cent;][]{1999ARA&A..37..363F, 2015ApJ...801...91D}. As mentioned earlier, the GBS-VLA results further showed that this emission mechanism can exhibit a broad range of spectral indices. The fact that only the YSOs at the phase centres were detected might suggest that either there are very few YSOs with active coronae in the TMC or that the non-thermal YSOs do not have very negative spectral indices as was shown by \citet{2015ApJ...801...91D}. However, the low angular resolution of our survey ($\sim10$\,arcsec) may also prevent detections of this non-thermal coronal emission which arises on much smaller scales close to the source.

Longer on-source time may yield larger YSO detection rates within these fields. To test this, we conducted a follow-up to the current pathfinder project with a blind survey of the crowded star forming region, NGC~1333 in the Perseus Molecular Cloud with the GMRT at 610\,MHz. These observations, which will be presented in a forthcoming paper, have a much longer integration time ($\sim58$\,hr) in order to take full advantage of the large field of view and showcase the potential survey capability of star forming regions with the GMRT. Furthermore, the capabilities of the GMRT are currently being enhanced with wider frequency coverage (almost seamless from 130 to 1500\,MHz), larger bandwidth (ten-fold increase from 32 to 400\,MHz) and more sensitive receivers. This upgraded GMRT (uGMRT), which has been designated as a SKA pathfinder, will be the most suited instrument for large-scale surveys of star forming regions at low radio frequencies.

\subsubsection{L1551~IRS~5 field}
\label{sec:L1551}

\begin{figure}
\includegraphics[width=\columnwidth]{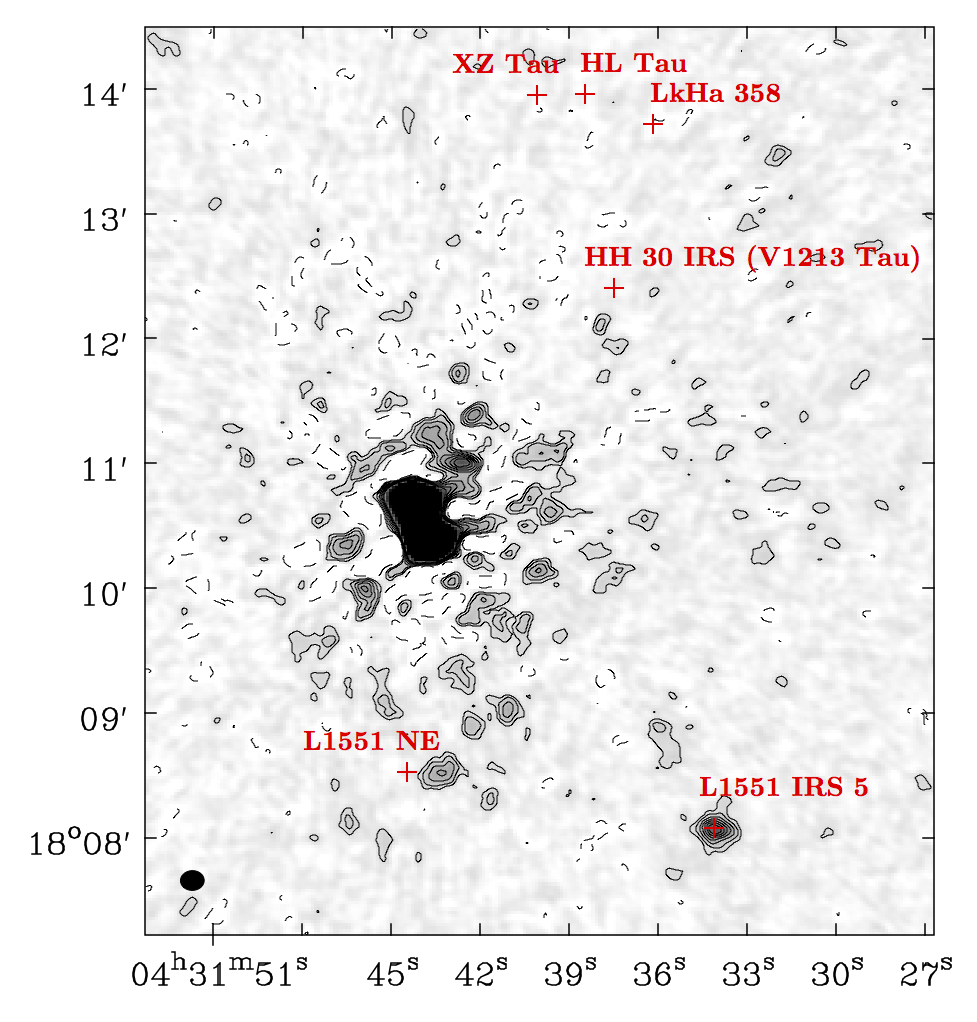}
\caption{The L1551 region at 608\,MHz. Greyscale ranges from $-0.1$ to 2\,mJy and contours are $-3$, 3, 5, 8, 11, 14, 17 and $20\times\sigma_{\rm rms}$ where $\sigma_{\rm rms}=54\,\umu$Jy\,beam$^{-1}$ (Table~\ref{tab:srclist}) is measured in a local patch of sky near L1551~IRS~5. We note that $\sigma_{\rm rms}$ will increase in the areas close to the bright radio galaxy dominating the map. Axes are J2000.0 coordinates and the \textsc{clean} restoring beam is shown as a filled ellipse in the bottom left corner (see Table~\ref{tab:srclist} for dimensions). Crosses denote locations of known YSOs. See Section~\ref{sec:L1551} for details.}
\label{fig:L1551region}
\end{figure}

The entire L1551 cloud itself fits within the FWHM of the GMRT primary beam at 608\,MHz \citep[and therefore also at 323\,MHz, see overview of L1551 region in H$\alpha$ and \rm{[S~\textsc{ii}]} in Fig.~1 of][]{1999AJ....118..972D}. Other well-studied YSO systems within this complex include L1551~NE, XZ~Tau, HL~Tau, LkH$\alpha$~358 and the driving source of HH~30 (V1213~Tau), the positions of which are shown in Fig.~\ref{fig:L1551region}. It can be seen that this region suffers from residual calibration artefacts due to a bright radio galaxy (northern lobe: GMRT-TAU~J043144.24+181041.8 with $S_{\rm 608\,MHz, int}=34.74\pm1.96$\,mJy and southern lobe: GMRT-Tau~J043143.78+181024.48 with $S_{\rm 608\,MHz, int}=60.00\pm3.14$\,mJy) which results in increased noise levels in this area of the map.

L1551~NE has been previously studied at centimetre wavelengths \citep[e.g.][]{2002AJ....124.1045R, 2012MNRAS.423.1089A}. Based on the spectral index and 4.5\,GHz flux density values from the GBS-VLA, L1551~NE (GBS-VLA~J043144.49+180831.6) has a predicted flux density of $0.32\pm0.45$\,mJy at 608\,MHz. Based on a local rms noise value of $93\,\umu$Jy\,beam$^{-1}$ (measured with the \textsc{AIPS} task \textsc{imean}) in the 608\,MHz (catalogue) map, we might expect to detect this object at $3\,\sigma_{\rm rms}$, however we do not and place a $3\,\sigma_{\rm rms}$ upper limit of $279\,\umu$Jy\,beam$^{-1}$ at 608\,MHz. There is a source of emission immediately to the west (peak emission located $\approx16$\,arcsec west) of the L1551~NE position, catalogued as GMRT-TAU~J043143.29+180831.5 with $S_{\rm 323\,MHz, int}=4.44\pm0.40$\,mJy and $S_{\rm 608\,MHz, int}=1.10\pm0.15$\,mJy. It is possible that this non-thermal emission is associated with the L1551~NE outflow similar to the non-thermal emission seen from the DG~Tau outflow \citep[][see Section~\ref{sec:dgtau} below]{2014ApJ...792L..18A}, however further observations are required to confirm this. 

XZ~Tau and HL~Tau are separated by $\sim25$\,arcsec, are both drivers of impressive outflows and are known radio emitters \citep[e.g.][]{1994ApJ...427L.103R, 2009ApJ...693L..86C}. The GBS-VLA calculates a spectral index of $\alpha_{\rm 4.5\,GHz}^{\rm 7.5\,GHz}=-0.85\pm0.36$ for XZ~Tau (GBS-VLA~J043140.09+181356.7), which is indicative of non-thermal (gyro)synchrotron radiation associated with magnetic activity. XZ~Tau therefore has a predicted flux density of $2.19\pm1.72$\,mJy at 608\,MHz based on the GBS-VLA data. This should be easily detectable within our survey, however from Fig.~\ref{fig:L1551region} we do not detect XZ~Tau at $3\,\sigma_{\rm rms}$ (where $58\,\umu$Jy\,beam$^{-1}$ is the local rms noise measured with \textsc{imean}). We therefore place an upper limit of $174\,\umu$Jy\,beam$^{-1}$ for XZ~Tau at 608\,MHz. We place the same upper limit on HL~Tau (GBS-VLA~J043138.42+181357.3) which has a predicted 608\,MHz flux density of $0.03\pm0.03$\,mJy based on GBS-VLA data. 

LkH$\alpha$~358 and V1213~Tau also drive outflows but were not detected by the GBS-VLA and have not previously been detected at centimetre wavelengths \citep[to the authors' knowledge, although LkH$\alpha$~358 was recently imaged with the Atacama Large Millimeter/submillimeter Array at 2.9\,mm, see][]{2015ApJ...808L...3A}, consistent with the non-detections in this work. We provide a $3\,\sigma_{\rm rms}$ upper limit of $174\,\umu$Jy\,beam$^{-1}$ for LkH$\alpha$~358 and V1213~Tau at 608\,MHz.

\subsubsection{DG~Tau field}
\label{sec:dgtau}

DG~Tau itself is not detected within the GMRT-TAU catalogue due to the fact that it is only detected at $3\,\sigma_{\rm rms}$ and the source fitting criteria requires a peak flux of $5\,\sigma_{\rm rms}$. We do however, detect the radio emission suggested to be a bow shock associated with the DG~Tau outflow \citep{2014ApJ...792L..18A} using the source finding criteria. The emission is detected as source GMRT-TAU~J042704.08+260559.3 with $S_{\rm 323\,MHz, int}=1.32\pm0.22$\,mJy, $S_{\rm 608\,MHz, int}=1.39\pm0.14$\,mJy and $\alpha_{\rm GMRT}=0.09\pm0.31$. These are not in agreement or within the errors of the results presented in \citet{2014ApJ...792L..18A} which could be due to the different methods of source fitting. \textsc{PyBDSM} fit this emission using a single Gaussian at 608\,MHz, despite its complicated structure, which may have given it an artificially high integrated flux density at this frequency. The peak flux densities of $S_{\rm 323\,MHz, peak}=0.90\pm0.15$\,mJy\,beam$^{-1}$, $S_{\rm 608\,MHz, peak}=0.50\pm0.12$\,mJy\,beam$^{-1}$ are in better agreement with the measurements and negative spectral index in \citet{2014ApJ...792L..18A}.

Located $\approx55''$ to the southwest of DG~Tau is DG~Tau~B, the driving source of the asymmetrical optical jet HH~$159$ \citep{1983ApJ...274L..83M} and not thought to be related to DG~Tau except by projected proximity \citep{1986ApJ...311L..23J}. DG~Tau~B has previously been detected at radio wavelengths \citep[e.g.][]{2012RMxAA..48..243R, 2012MNRAS.420.3334S, 2013MNRAS.436L..64A} and, based on the GBS-VLA data (GBS-VLA~J042702.56+260530.4), has a predicted 608\,MHz flux density of $0.12\pm0.08$\,mJy. With a local rms noise of $80\,\umu$Jy\,beam$^{-1}$, we place a $3\,\sigma_{\rm rms}$ upper limit of $240\,\umu$Jy\,beam$^{-1}$ for DG~Tau~B at 608\,MHz.

\subsection{Comparison between 323 and 608\,MHz images}

\begin{figure*}
\subfloat[323\,MHz Peak Flux Density]{\includegraphics[width=0.5 \textwidth]{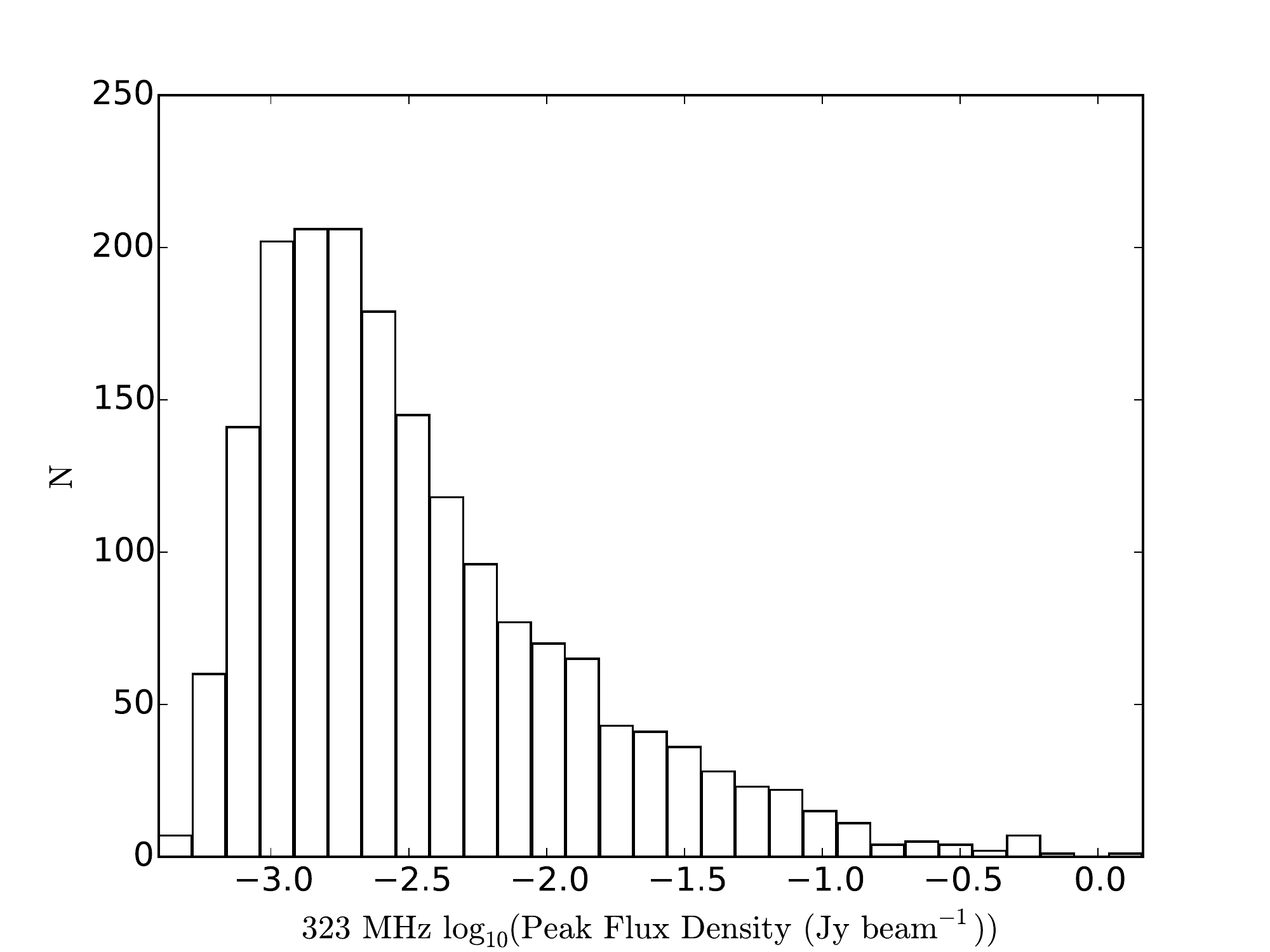} \label{fig:325peak}}
\subfloat[608\,MHz Peak Flux Density]{\includegraphics[width=0.5 \textwidth]{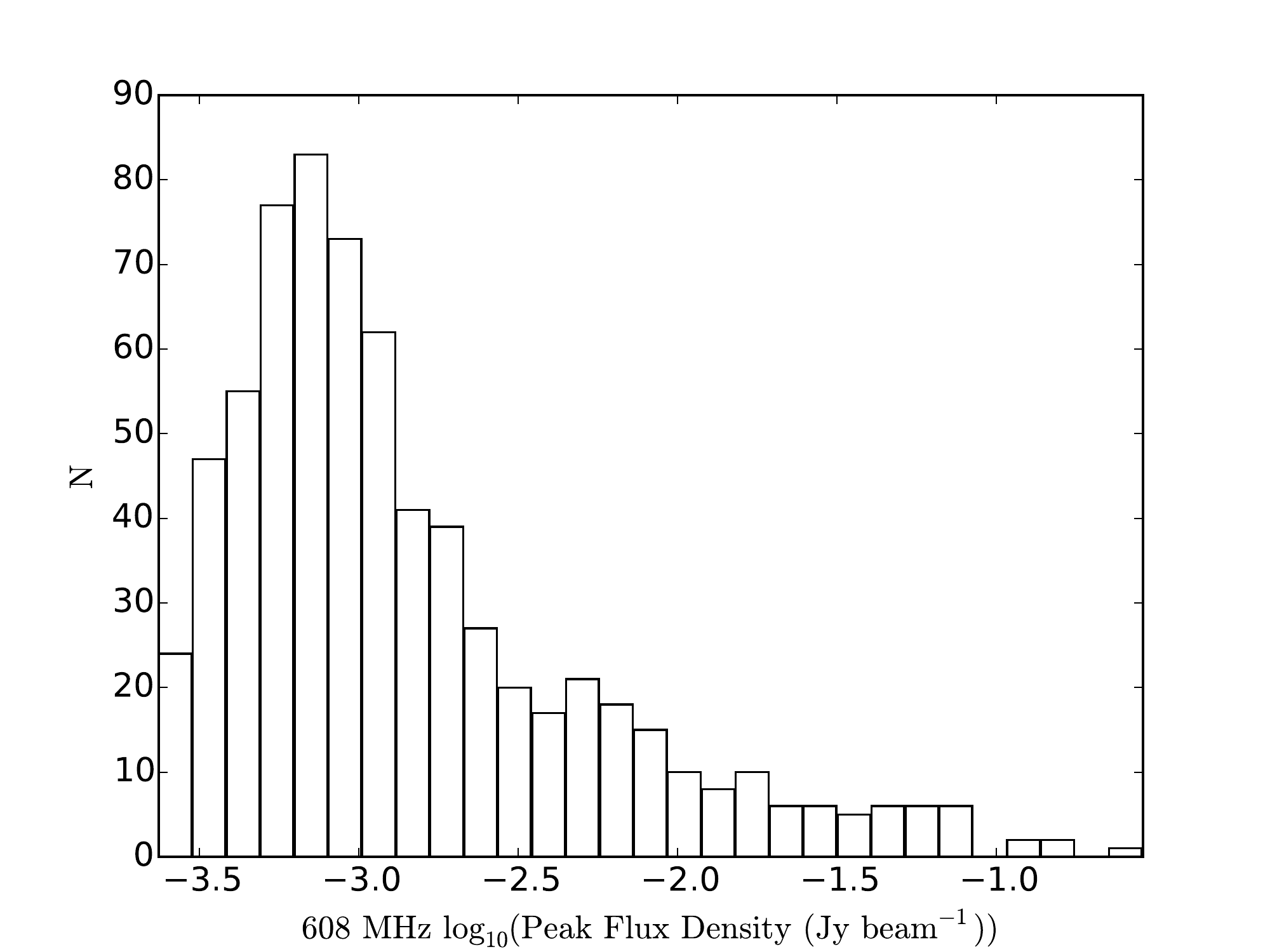} \label{fig:610peak}} \\
\subfloat[323\,MHz Integrated Flux Density]{\includegraphics[width=0.5 \textwidth]{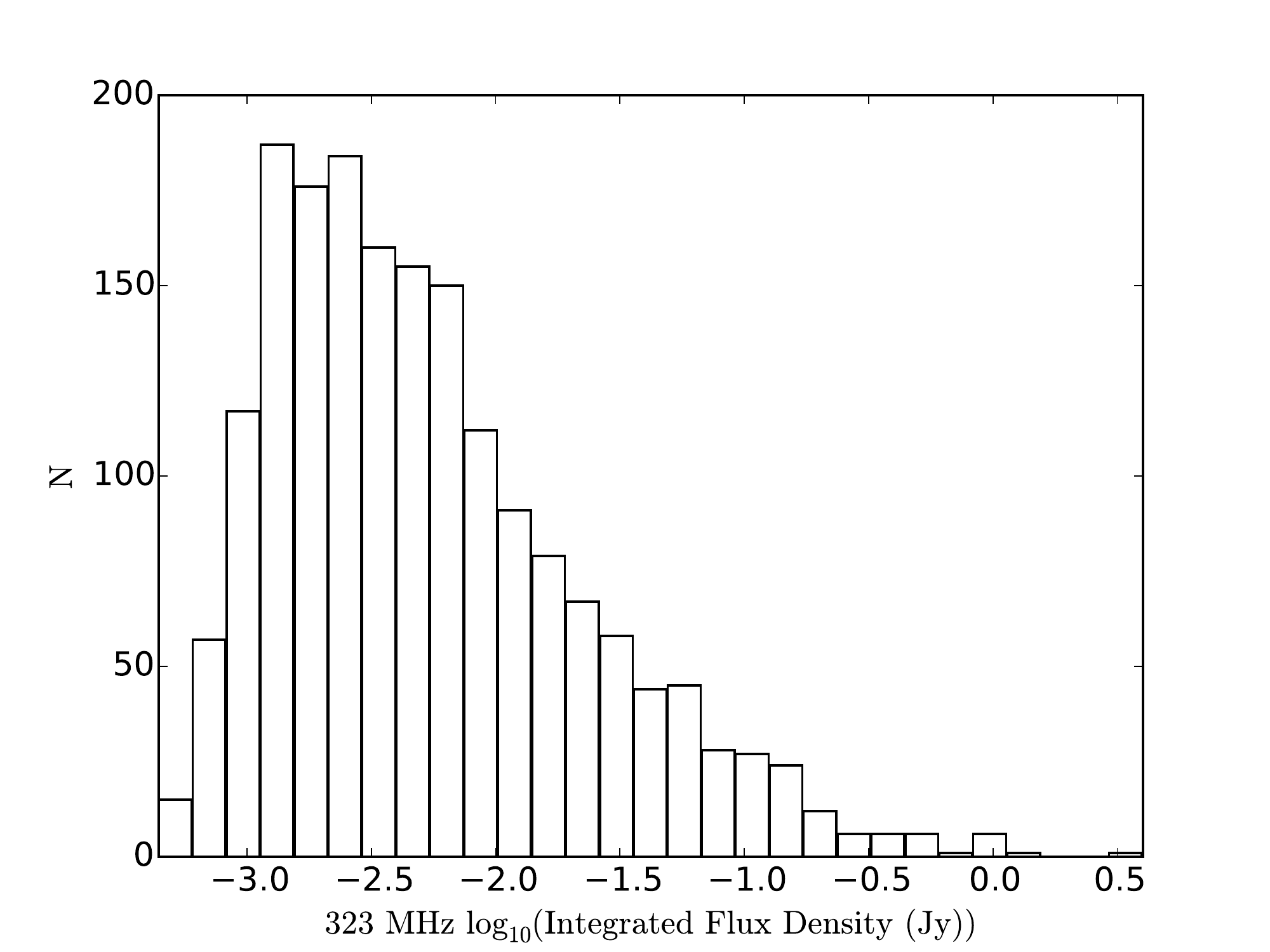}  \label{fig:325int}} 
\subfloat[608\,MHz Integrated Flux Density]{\includegraphics[width=0.5 \textwidth]{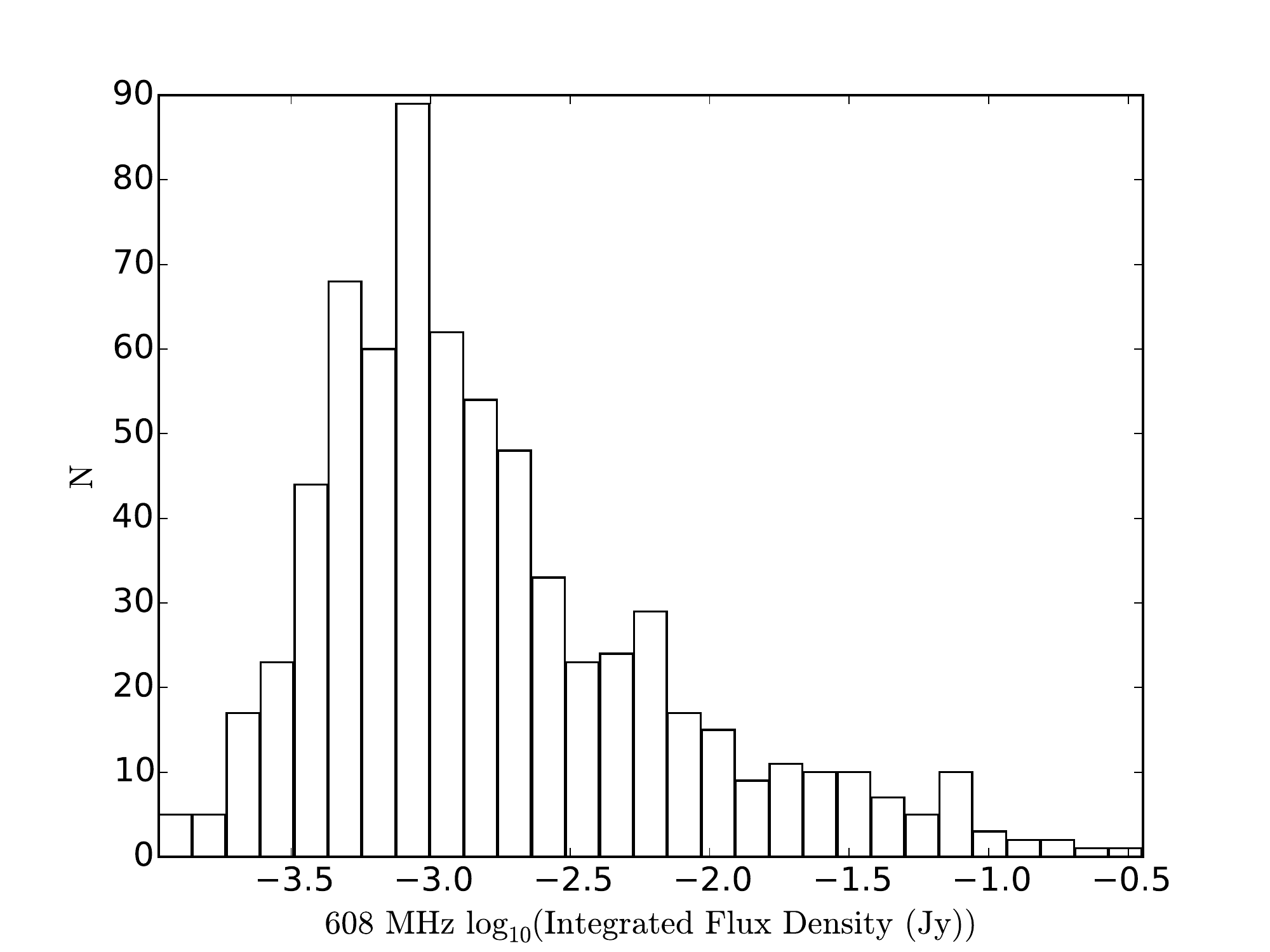} \label{fig:610int}} \\
\subfloat[Uncertainty in 323\,MHz Peak Flux Density]{\includegraphics[width=0.5 \textwidth]{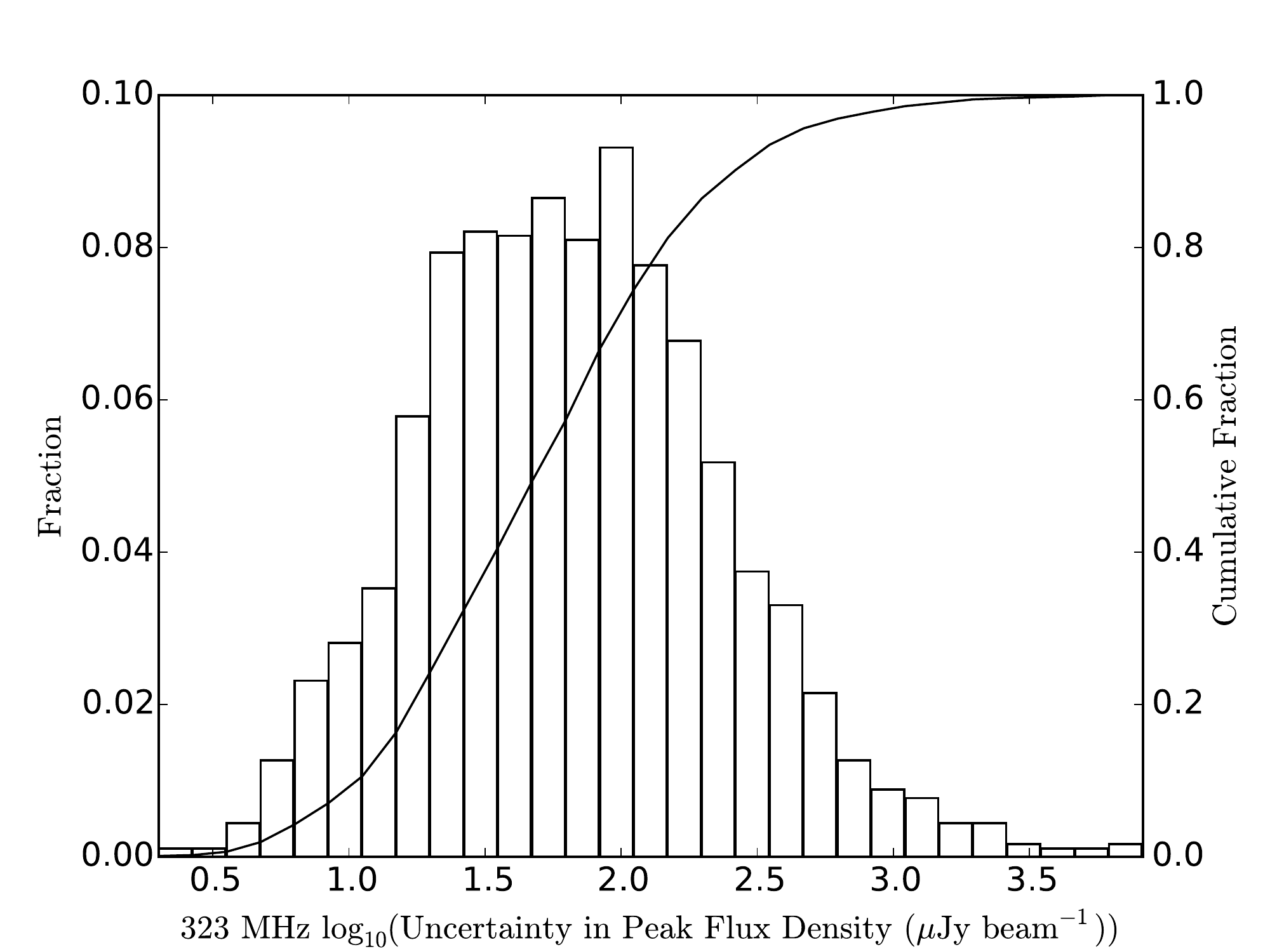} \label{fig:325uncertainty}}
\subfloat[Uncertainty in 608\,MHz Peak Flux Density]{\includegraphics[width=0.5 \textwidth]{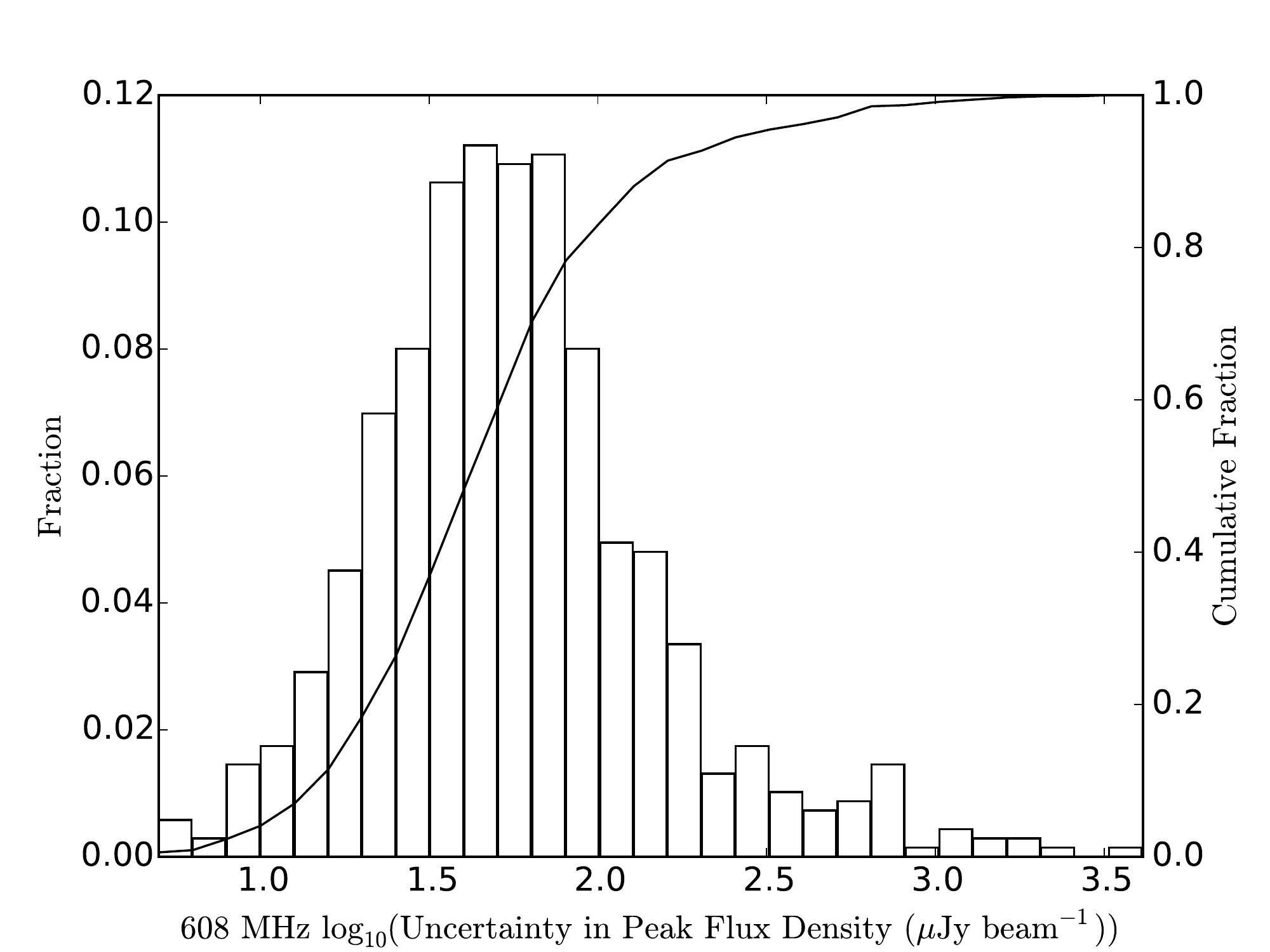} \label{fig:610uncertainty}}
\caption[Flux density distributions.]{Flux density distribution across the catalogue sources at 323 and 608\,MHz. Each row of panels present the peak flux density (a and b), integrated flux density (c and d) and \textsc{PyBDSM} peak flux density uncertainty (e and f) at 323 and 608\,MHz. The continuous line in (e) and (f) indicates the cumulative distribution function.}
\label{fig:cat}
\end{figure*}

Fig.~\ref{fig:cat} shows the distribution of peak flux density, integrated flux density and the \textsc{Resid\_Isl\_rms} flux (local rms noise) across the survey at 323 and 608\,MHz. The residual island rms flux is the rms of the flux left in an island after the modelled source(s) on the island have been subtracted, and is thus a combination of a measure of the local noise and the quality of the fit.

\begin{figure*}
\subfloat[323 to 608 MHz Spectral Index]{\includegraphics[width=0.5 \textwidth]{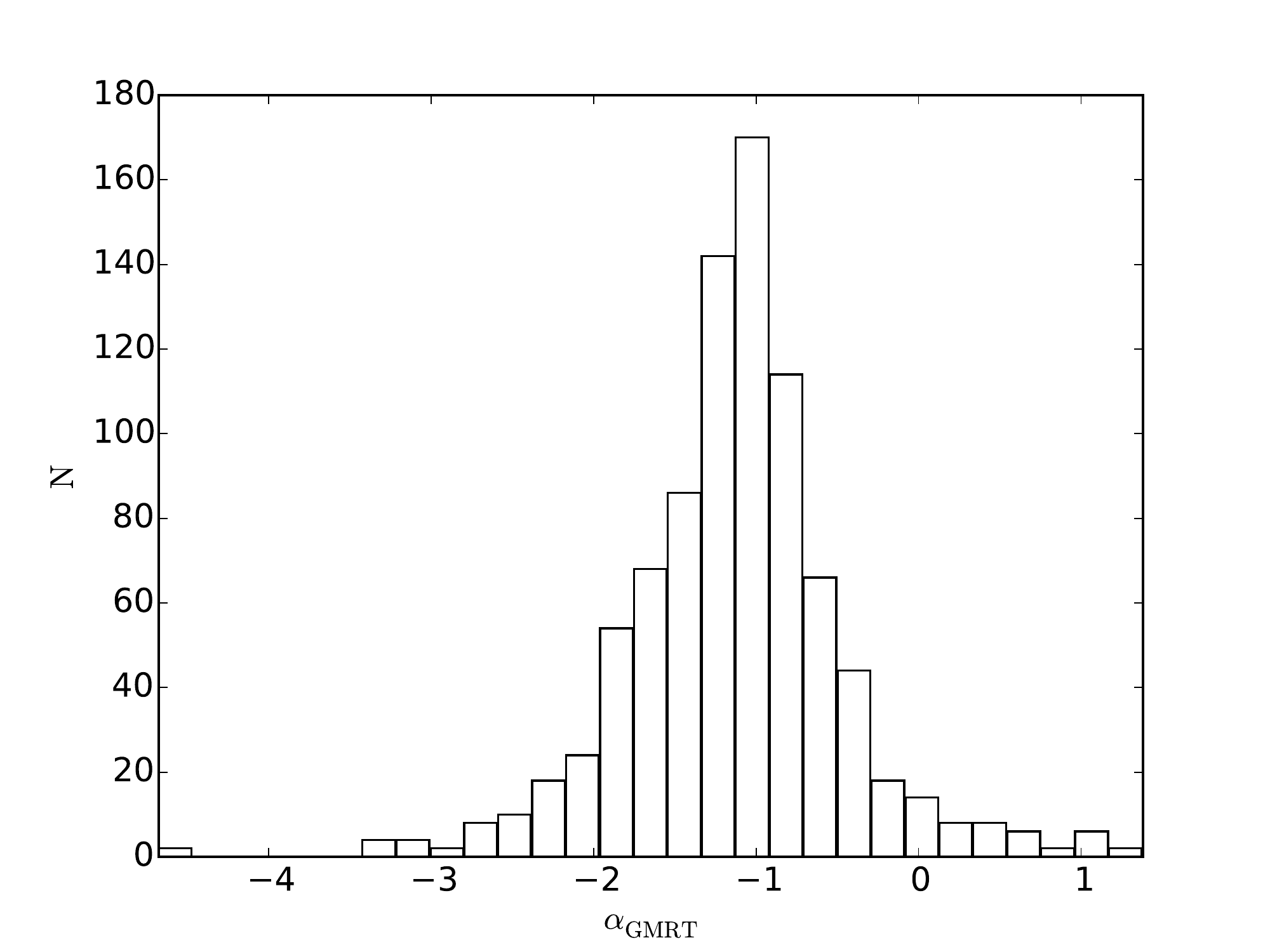} \label{fig:gmrt_spindx}}
\subfloat[Distribution of 323 to 608 MHz Spectral Index to Flux Density]{\includegraphics[width=0.5 \textwidth]{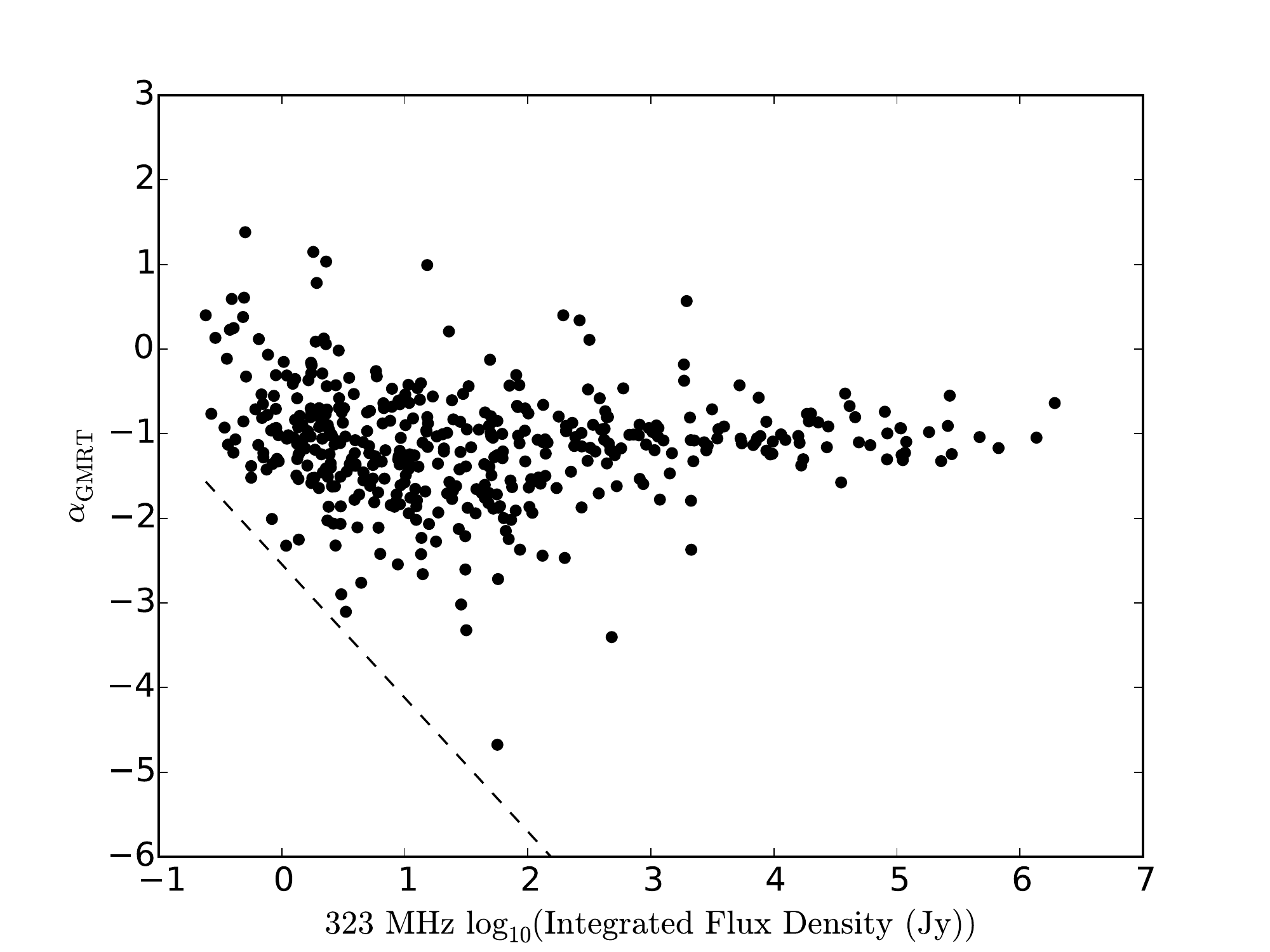} \label{fig:gmrt_spindx_dist}} 
\caption[Spectral index distributions.]{(a) The 323 to 608\,MHz GMRT spectral index distribution with a median value of $\alpha_{\rm GMRT}=-1.10$ and a standard deviation of $0.67$. (b) Plot of $\alpha_{\rm GMRT}$ vs. the integrated flux density at 323\,MHz. The dashed line indicates the $3\,\sigma_{\rm rms}$, $200\,\umu$Jy\,beam$^{-1}$ sensitivity threshold at which the higher frequency (608\,MHz) would not be able to detect (non-thermal) emission.}
\end{figure*}

Fig.~\ref{fig:gmrt_spindx} shows the distribution of $\alpha_{\rm GMRT}$ which has a median of $-1.10$ and a standard deviation of $0.67$, consistent with a majority of sources in the survey being Active Galactic Nuclei (AGN) emitting synchrotron radiation \citep[see e.g.][]{1970ranp.book.....P}. Fig.~\ref{fig:gmrt_spindx_dist} shows the variation of spectral index with the log of the integrated source flux. As expected, fainter sources with lower signal to noise ratios result in a larger spread of spectral index values. A notable observational bias can be seen at the lowest fluxes, where faint non-thermal sources detected at 323\,MHz are unlikely to be detected at 608\,MHz, thus leaving an apparent gap in the bottom left of the plot. The dashed line plotted shows the range of 323\,MHz fluxes and spectral index values corresponding to the $3\,\sigma_{\rm rms}$ sensitivity at 608\,MHz of $\sim200\,\umu$Jy\,beam$^{-1}$. No detections are expected below this line.

\begin{figure*}
\subfloat[GMRT-TAU J043051.62+180819.1 and GMRT-TAU J043053.60+180752.4]{\includegraphics[width=0.29\textwidth]{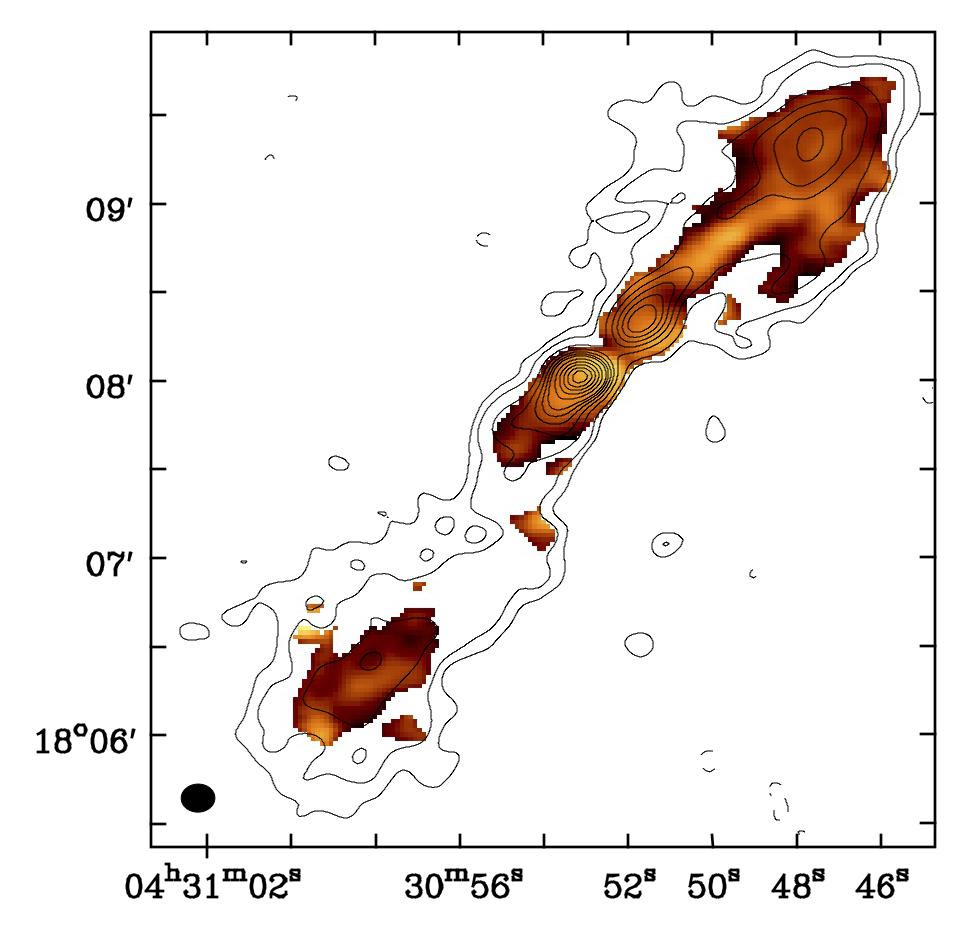}} \qquad
\subfloat[GMRT-TAU J043140.25+180325.8 and GMRT-TAU J043141.23+180248.3]{\includegraphics[width=0.26\textwidth]{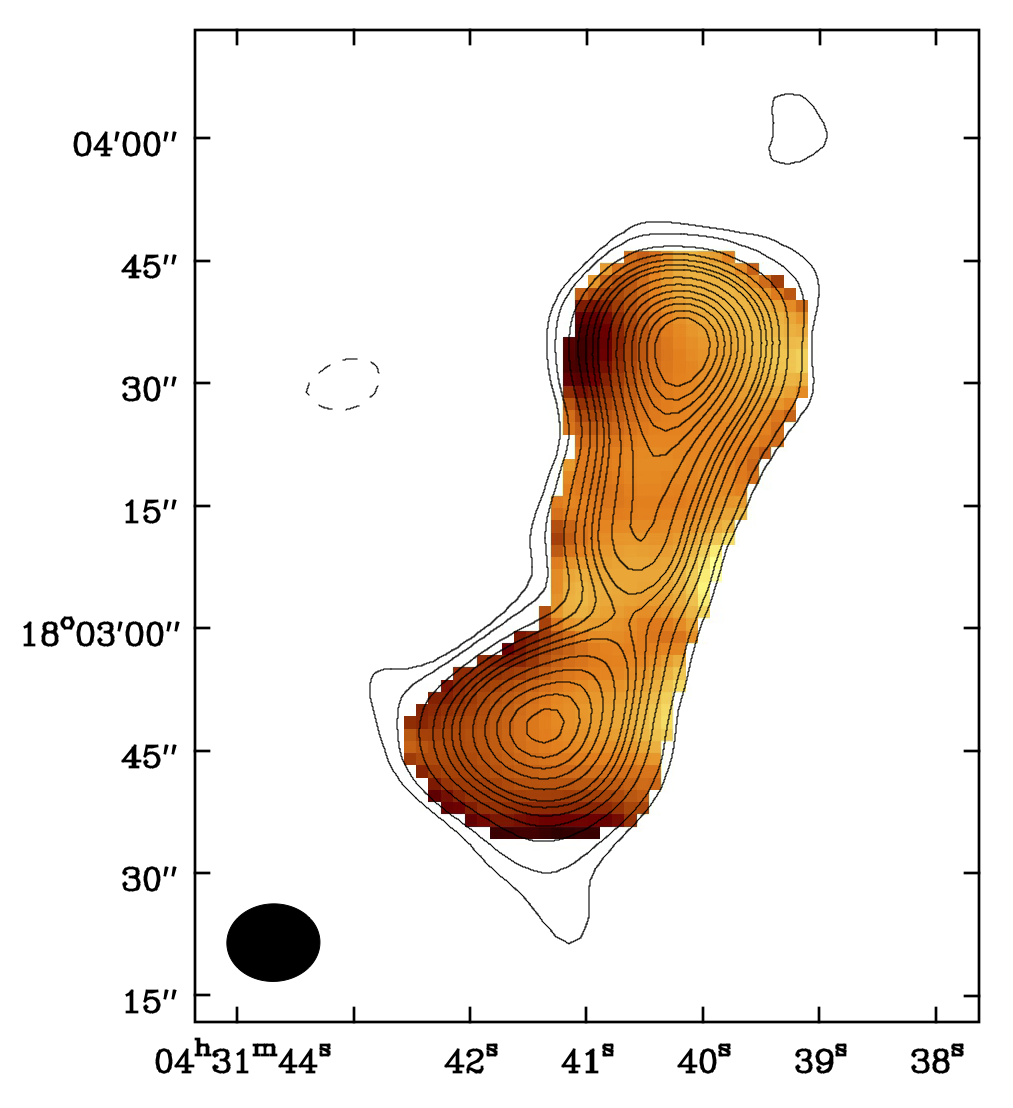}} \qquad
\subfloat[GMRT-TAU J042729.21+255046.0, GMRT-TAU J042730.16+255121.0 and GMRT-TAU J042730.32+255013.2]{\includegraphics[width=0.35\textwidth]{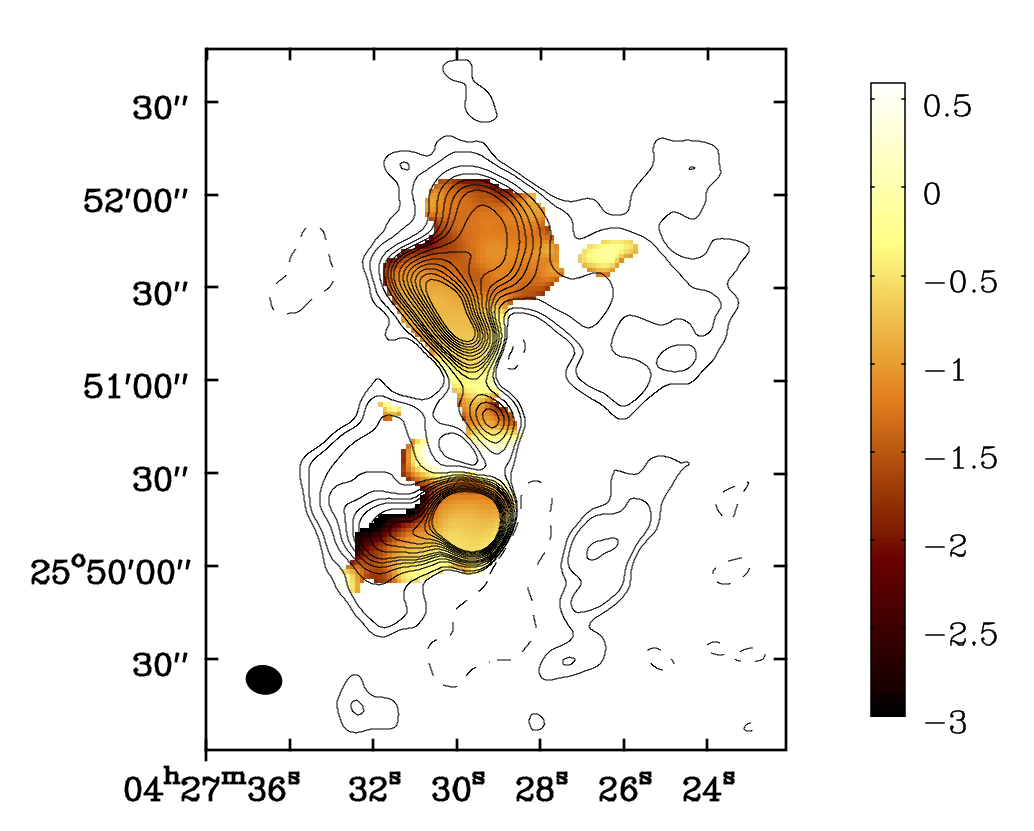}}
\caption{A sample of spectral index maps for extended objects in the GMRT-TAU fields. (a) and (b) are located within the L1551~IRS~5 field, and (c) is located within the DG~Tau field. Greyscale (colour in the online journal) represents the range in spectral index. Contours are plotted as $-3$, 3, 5, 10, 15, 20, 25, 30, 40, 50, 60, 70, 80, 90 and $100\times\sigma_{\rm rms}$, where $\sigma_{\rm rms}$ is the local rms noise in the 323\,MHz image and equals (a) 158, (b) 148 and (c) $142\,\umu$Jy\,beam$^{-1}$. Axes are J2000.0 coordinates and the \textsc{clean} restoring beam is shown as a filled ellipse in the bottom left corner of each map. Fluxes were clipped below $5\,\sigma_{\rm rms}$ at both frequencies during spectral index map construction to eliminate artificially steep indices caused by low SNR emission, however we show $3\,\sigma_{\rm rms}$ contours to remain consistent with the source fitting procedure (Section~\ref{sec:sf}) and Fig.~\ref{fig:maps}.}
\label{fig:spix}
\end{figure*}

In Fig.~\ref{fig:spix} we present spectral index maps of three extended radio galaxies detected in our survey made using the \textsc{spix} operation within the AIPS task \textsc{comb}. We set \textsc{aparm(9)} and \textsc{(10)} to be the local $5\,\sigma_{\rm rms}$ cutoff for each frequency to eliminate low signal-to-noise (SNR) emission which can create artificially steep spectral indices. In general, the bright regions coincide with flatter indices ($\alpha_{\rm GMRT}\approx-1$) and the more diffuse regions exhibit steeper indices ($\alpha_{\rm GMRT}\approx-2$). This steeping can be interpreted in terms of spectral ageing of electrons as they move from acceleration sites in the bright radio knots to regions further along the jet, by which time synchrotron cooling has reduced the population of high energy electrons relative to the lower energies \citep{1962SvA.....6..317K}. Therefore the steepening trend observed is realistic, however we caution that the spectral indices themselves may not be due to the relatively high errors involved in a two point fit of measurements close in frequency.

\subsection{Comparison with NVSS}

\begin{figure*}
\subfloat[GMRT to NVSS Spectral index]{\includegraphics[width=0.5 \textwidth]{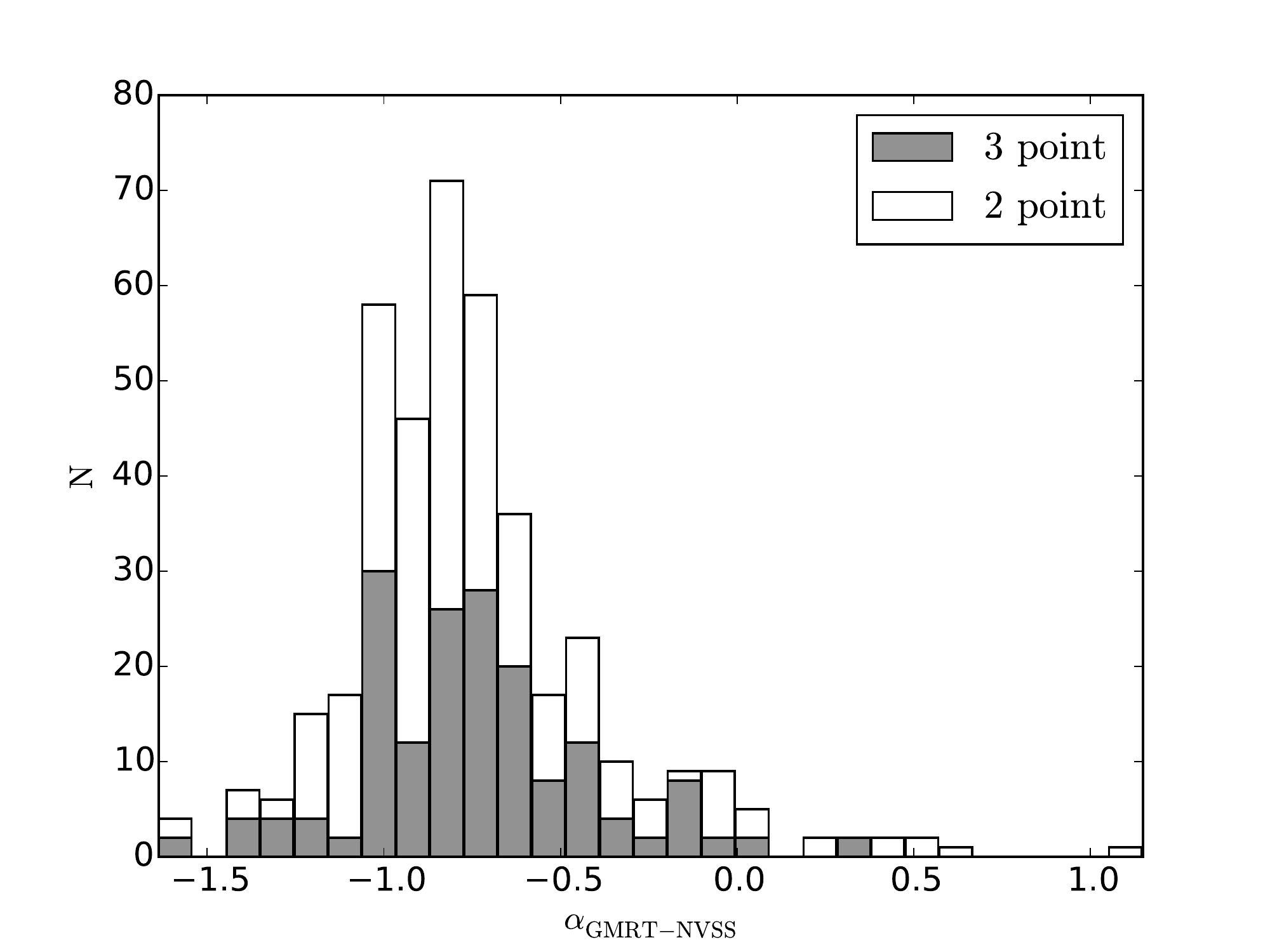} \label{fig:nvss_spindx}} 
\subfloat[Distribution of GMRT to NVSS Spectral Index to Flux Density]{\includegraphics[width=0.5 \textwidth]{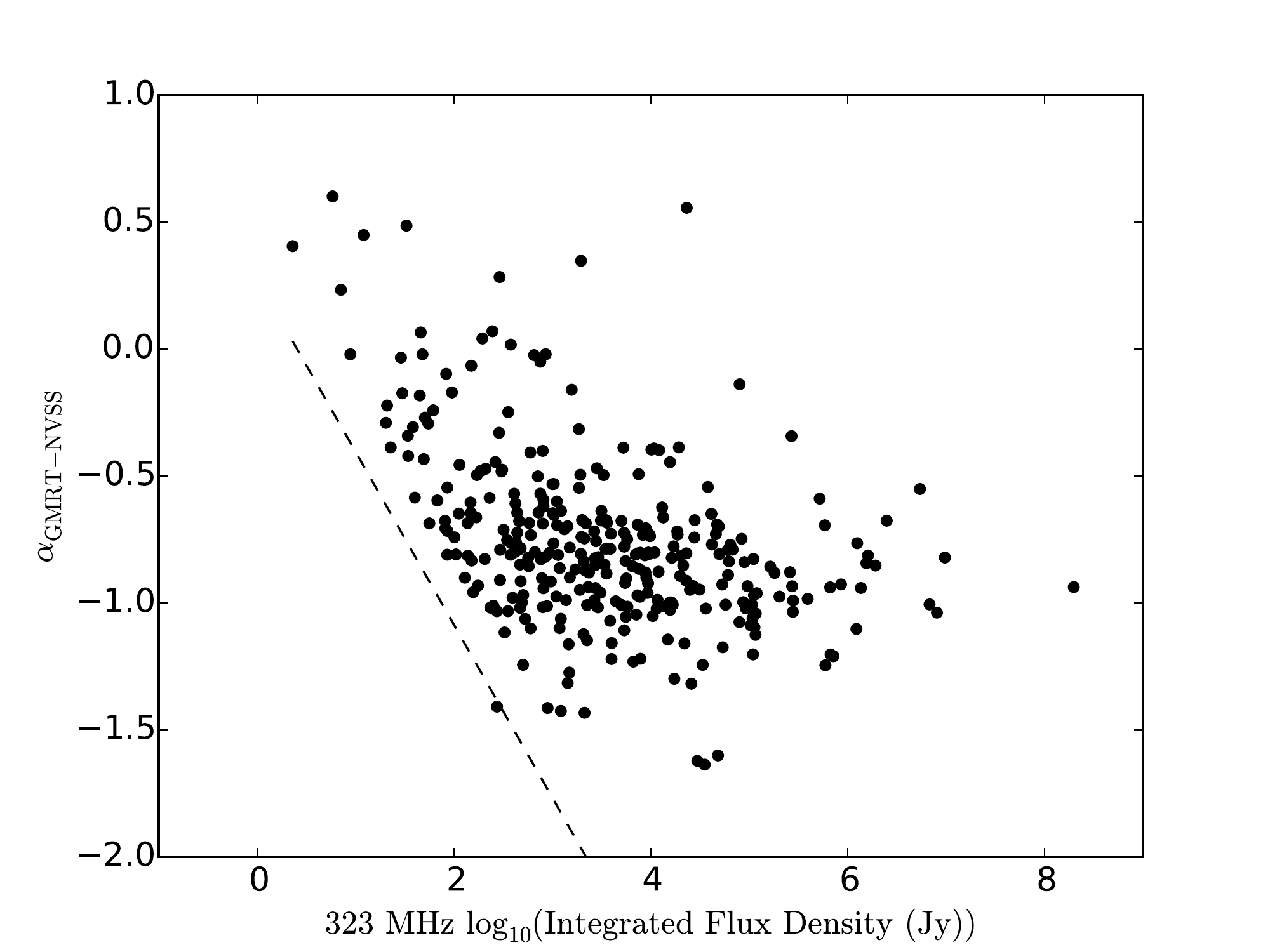} \label{fig:nvss_spindx_dist}} 
\caption[Spectral index distributions.]{ (a) The GMRT to NVSS (1.4\,GHz) spectral index distribution which has a median value of $\alpha_{\rm NVSS}=-0.80$ with a standard deviation of $0.36$. (b) Plot of $\alpha_{\rm NVSS}$ vs. the integrated flux density at 323\,MHz. The dashed line indicates the $3\,\sigma_{\rm rms}$, 1.5\,mJy\,beam$^{-1}$ sensitivity threshold at which the higher frequency (1.4\,GHz VLA) would not be able to detect (non-thermal) emission.}
\end{figure*}

Fig.~\ref{fig:nvss_spindx} shows the distribution of $\alpha_{\rm NVSS}$ which has a median of $-0.80$ and a standard deviation of $0.36$, again consistent with synchrotron radiation from AGN. Note that these plots present a combination of two and three frequency calculations of the spectral index, depending on whether the source was detected at both GMRT frequencies or not. Fig.~\ref{fig:nvss_spindx_dist} again shows a gap in the bottom left of the plot due to observational bias, this time exaggerated due to the larger spread in frequencies. The dashed line plotted shows the range of 323\,MHz fluxes and spectral index values corresponding to the $3\,\sigma_{\rm rms}$ sensitivity at 1.4\,GHz of $\sim1.5$\,mJy\,beam$^{-1}$. No detections are expected below this line.

As described in Section~\ref{sec:obs}, a small systematic offset appeared to be present between the 323 and 608\,MHz maps of the L1551~IRS~5 and DG~Tau fields. The positions of the fitted bright, compact sources in each field at 608\,MHz are in general consistent with those from the NVSS to within an average of 0.5\,arcsec, making them more reliable than the positions at 323\,MHz. However, due to a lack of low frequency radio surveys at higher resolution for this patch of the sky (NVSS has a spatial resolution of approximately 45\,arcsec) the absolute coordinates may have a residual uncertainty of around 0.5\,arcsec. This should not have a significant impact as the GMRT spatial resolution is of order 5--10\,arcsec and we do not apply further correction as it is uncertain as to which survey the errors come from \citep[see e.g.][]{2007MNRAS.376.1251G}.

\vspace{-10pt}
\section{Data Products}
\label{sec:dp}

The data products which are released as part of this GMRT survey include images, the source catalogue and the scripts which were run to create the source catalogue and validation plots. These data products are hosted on the project website\footnote[3]{https://homepages.dias.ie/rainsworth/GMRT-TAU\_catalogue.html} and are described below. \\

\subsection{Images} 

We provide image files in FITS format for each field at each frequency at both native and re-imaged resolutions with primary beam correction applied (12 FITS images in total). The following files were used to create the source catalogue presented in this work (see Section~\ref{sec:obs}), each with an image cell size of 1.5\,arcsec:
\begin{description}
\item [\textbf{L1551\_323MHz\_catalogue.fits}] L1551~IRS~5 field at 323\,MHz imaged with a native resolution ($11.4\times9.5$\,arcsec$^2$, $-88.5^\circ$).
\item [\textbf{L1551\_608MHz\_catalogue.fits}] L1551~IRS~5 field at 608\,MHz convolved with the 323\,MHz resolution ($11.4\times9.5$\,arcsec$^2$, $-88.5^\circ$).
\item [\textbf{TTau\_323MHz\_catalogue.fits}] T~Tau field at 323\,MHz imaged with a native resolution ($10.8\times9.5$\,arcsec$^2$, $-81.6^\circ$).
\item [\textbf{TTau\_608MHz\_catalogue.fits}] T~Tau field at 608\,MHz convolved with the 323\,MHz resolution ($10.8\times9.5$\,arcsec$^2$, $-81.6^\circ$).
\item [\textbf{DGTau\_323MHz\_catalogue.fits}] DG~Tau field at 323\,MHz imaged with a native resolution ($11.6\times9.2$\,arcsec$^2$, $79.6^\circ$).
\item [\textbf{DGTau\_608MHz\_catalogue.fits}] DG~Tau field at 608\,MHz convolved with the 323\,MHz resolution ($11.6\times9.2$\,arcsec$^2$, $79.6^\circ$).
\end{description}

\noindent The following files are the original image files used to detect the target sources, the results of which were presented in \citet{2016MNRAS.459.1248A}:
\begin{description}
\item [\textbf{L1551\_323MHz\_native.fits}] L1551~IRS~5 field at 323\,MHz imaged with a native resolution ($11.4\times9.5$\,arcsec$^2$, $-88.5^\circ$) and 3\,arcsec cell size.
\item [\textbf{L1551\_608MHz\_native.fits}] L1551~IRS~5 field at 608\,MHz imaged with a native resolution ($6.2\times4.9$\,arcsec$^2$, $76.5^\circ$) and 1.5\,arcsec cell size.
\item [\textbf{TTau\_323MHz\_native.fits}] T~Tau field at 323\,MHz imaged with a native resolution ($10.8\times9.5$\,arcsec$^2$, $-81.6^\circ$) and 3\,arcsec cell size.
\item [\textbf{TTau\_608MHz\_native.fits}] T~Tau field at 608\,MHz imaged with a native resolution ($6.0\times5.0$\,arcsec$^2$, $83.8^\circ$) and 1.5\,arcsec cell size.
\item [\textbf{DGTau\_323MHz\_native.fits}] DG~Tau field at 323\,MHz imaged with a native resolution ($11.6\times9.2$\,arcsec$^2$, $79.6^\circ$) and 3\,arcsec cell size.
\item [\textbf{DGTau\_608MHz\_native.fits}] DG~Tau field at 608\,MHz imaged with a native resolution ($6.5\times5.2$\,arcsec$^2$, $74.0^\circ$) and 1.5\,arcsec cell size.
\end{description}

\subsection{Catalogue and scripts} 

The final source catalogue as described in Section~\ref{sec:cat_creation} can be found in the Supplementary Material through the online version of this article. It will also be made available in a machine-readable format as part of the data release on both the project website\footnotemark[3] and VizieR online database. We note that the columns will differ slightly in the machine-readable format than described in Section~\ref{sec:cat_creation} due to additional column requirements for errors, therefore a ReadMe file will be included to describe column headings. In addition, a table listing all the parameter outputs from \textsc{PyBDSM} will be provided, see the online documentation\footnotemark[2] for a full description. 

A series of \textsc{python} scripts were developed to create the final catalogue and validation plots from the \textsc{PyBDSM} catalogue files for each individual field and frequency. All of these files are available on the project website and can be used to re-constitute the catalogue at any time. \textsc{make\_catalog.py} takes the \textsc{PyBDSM} output for a single field at both frequencies and generates a final catalogue,  complete with spectral indices, source matching between frequencies, and comparisons with other surveys. It outputs these results in a \LaTeX\ format that can then be combined with results from additional fields to make the final GMRT-TAU catalogue. Additional files are also created which can be read by \textsc{make\_plots.py} to generate the diagnostic plots presented in this paper. Finally, a .csv file is created which contains all of the information in the final catalogue, in addition to default \textsc{PyBDSM} columns and additional columns used for internal calculation by \textsc{make\_catalog.py}, but which may be of use in further study.

\vspace{-10pt}
\section{Summary}
\label{sec:summary}

In this paper we have described a 323 and 608\,MHz (90 and 50\,cm, respectively) survey with the Giant Metrewave Radio Telescope of three regions towards the Taurus Molecular Cloud, specifically towards the target pre-main-sequence stars L1551~IRS~5, T~Tau and DG~Tau. This survey is a natural by-product of the large instantaneous field of view of the GMRT. Although we did not detect other YSOs in addition to the targets based on our source finding criteria, we provide a catalogue of field sources which can be useful for targeted manual searches, studies of radio galaxies or to assist in the calibration of future observations with LOFAR towards these regions. 

We emphasise that the work presented here is part of a pathfinder project to detect the specific YSO target sources at the phase centre of each pointing, which was successful \citep{2016MNRAS.459.1248A}. Longer on-source time may yield larger YSO detection rates within these fields and we have conducted a follow-up survey of the crowded star forming region NGC~1333 in the Perseus Molecular Cloud with a much longer integration time at 610\,MHz (Ainsworth et~al., in preparation). These observations will showcase the potential survey capability of star forming regions with the GMRT. 

The resolution of the survey is of order 10\,arcsec and the best rms noise at the centre of each pointing is of order $100\,\umu$Jy\,beam$^{-1}$ at 323\,MHz and $50\,\umu$Jy\,beam$^{-1}$ at 608\,MHz. The final data products comprise 12 images, the final source catalogue, an additional catalogue containing the results corresponding to all \textsc{PyBDSM} parameters and the scripts used to generate the catalogue and validation plots. These data will be made available on the project website\footnotemark[3] and on the VizieR online database. 

The final catalogue contains 1815 sources at 323\,MHz and 687 sources at 608\,MHz. A total of 440 sources were detected at both frequencies which yields a total unique source count of 2062. The catalogue has been cross-referenced with existing radio, infrared and X-ray surveys conducted towards this region and compared with other GMRT surveys and the NVSS. Notable YSO non-detections have been discussed and a sample of extended radio galaxies has been shown.  

The capabilities of the GMRT are currently being enhanced with wider frequency coverage, larger bandwidth and more sensitive receivers. This uGMRT has been designated as a SKA pathfinder and will be the most suited instrument for large-scale surveys of star forming regions at very low radio frequencies.

\vspace{-10pt}
\section*{Acknowledgements}

We thank the referee, Laurent Loinard, for his constructive comments which helped to clarify these results. We thank the staff of the GMRT who have made these observations possible. GMRT is run by the National Centre for Radio Astrophysics of the Tata Institute of Fundamental Research. REA, CPC and TPR would like to acknowledge support from Science Foundation Ireland under grant 13/ERC/I2907. DAG thanks the Science and Technology Facilities Council for support. AMS gratefully acknowledges support from the European Research Council under grant ERC-2012-StG-307215 LODESTONE.




\bibliographystyle{mnras}
\bibliography{Bibliography} 



\section*{Supplementary Material}

The following supplementary material is available for this article: \\

\noindent \textbf{Table 3.} The full version of the 323 and 608\,MHz GMRT-TAU catalogue, sorted by right ascension.


\bsp	
\label{lastpage}
\end{document}